\documentclass[manuscript,screen]{acmart}
\AtBeginDocument{%
  }

\setcopyright{none}
\usepackage{enumitem}
\usepackage{graphicx}
\usepackage{booktabs}
\usepackage{tabularx}
\usepackage{array}
\usepackage{soul}
\usepackage{longtable}  
\usepackage{booktabs}
\usepackage{array}
\newcolumntype{Y}{>{\raggedright\arraybackslash}X}
\newcommand{\pid}[1]{\textbf{\textsf{#1}}}

\begin{document}

\title{Workplace Ambiguity and Technology Support for Autistic and IDD Workers}

\author{Silas Curry}
\email{currysc@mail.uc.edu}
\affiliation{%
  \institution{University of Cincinnati}
  \city{Cincinnati}
  \state{Ohio}
  \country{United States}
}

\author{Rishit Singh}
\email{singh3rt@mail.uc.edu}
\affiliation{%
  \institution{University of Cincinnati}
  \city{Cincinnati}
  \state{Ohio}
  \country{United States}
}

\author{Heekyoung Jung}
\email{junghg@ucmail.uc.edu}
\affiliation{%
  \institution{University of Cincinnati}
  \city{Cincinnati}
  \state{Ohio}
  \country{United States}
}

\author{Angel Hsing-Chi Hwang}
\email{angel.hwang@usc.edu}
\affiliation{%
  \institution{University of Southern California}
  \city{Los Angeles}
  \state{California}
  \country{United States}
}

\author{Fan Yang}
\email{YANG259@mailbox.sc.edu}
\affiliation{%
  \institution{University of South Carolina}
  \city{Columbia}
  \state{South Carolina}
  \country{United States}
}

\author{Ce Zhong}
\email{ce.zhong@asu.edu}
\affiliation{%
  \institution{Arizona State University}
  \city{Tempe}
  \state{Arizona}
  \country{United States}
}

\author{Nicholas Lindgren}
\email{nicholas.lindgren@uc.edu}
\affiliation{%
  \institution{University of Cincinnati}
  \city{Cincinnati}
  \state{Ohio}
  \country{United States}
}

\author{Chen Chen}
\email{chechen@fiu.edu}
\affiliation{%
  \institution{Florida International University}
  \city{Miami}
  \state{Florida}
  \country{United States}
}

\author{Lingyao Li}
\email{lingyaoli@arizona.edu}
\affiliation{%
  \institution{College of Information Science, University of Arizona}
  \city{Tucson}
  \state{Arizona}
  \country{United States}
}

\author{Renkai Ma}
\email{mark@ucmail.uc.edu}
\affiliation{%
  \institution{University of Cincinnati}
  \city{Cincinnati}
  \state{Ohio}
  \country{United States}
}
\renewcommand{\shortauthors}{Anonymous Author(s)}

\begin{abstract}
Workplace ambiguity, where expectations, evaluations, or procedures are left unstated, acts as a structural access barrier for autistic and IDD workers, routinely forcing them to act on guesses. To investigate how these barriers operate, we interviewed 11 caregivers, concluding with an AI co-design activity to explore technological solutions. Their perspectives revealed five themes of workplace ambiguity, which we synthesize into a conceptual framework comprising three intersecting dimensions: evaluative, procedural, and relational. To address these barriers, participants proposed three forms of technology support, including ambient real-time redirection, feedback processing, and physiological guidance, designed to make hidden expectations perceptible and actionable. However, this exposes a core tension: technologies designed to make the workplace legible can inadvertently make the worker hyper-legible to employers through surveillance. We argue that ambiguity-aware workplaces must make expectations explicit before they carry consequences, and assistive technologies must remain accountable to the worker.
\end{abstract}



\maketitle
\section{Introduction}
Autism is a neurodevelopmental condition involving persistent differences in social communication and interaction alongside restricted or repetitive patterns of behavior, interests, or sensory responses \cite{lord2020autism}. The Centers for Disease Control and Prevention identified it in one of every 31 eight-year-olds across its 16 surveillance sites in the US in 2022 \cite{shaw2025prevalence}. Intellectual and developmental disabilities (IDD) originate during the developmental period and involve limitations in intellectual functioning, adaptive behavior, or both \cite{schalock2021overview}. Among young adults who had received special education services, autistic adults had the lowest rate of paid employment among the disability categories \cite{roux2013postsecondary}, and among adults with IDD who receive services from state developmental disabilities agencies, 17\% hold a paid job in the community while 45\% of those without one report wanting one \cite{nci2026employment}. The two are distinct conditions, but we study these workers together and refer to them throughout as \emph{autistic and IDD workers}, because an expectation a workplace never states can affect clinically different workers in much the same way. Employment research groups them for a related reason, since both meet overlapping structural barriers to competitive work \cite{clear2022supported,hedley2017employment}.

Getting hired is only the first barrier autistic and IDD people face at work. Close to a third of 315 autistic adults in a study reported having been terminated from a job \cite{pezzimenti2024termination}, and a twelve-year study of autistic adults without intellectual disability found that fewer than a quarter sustained competitive employment or postsecondary education across the period \cite{taylor2015longitudinal}. Current interventions have primarily focused on modifying autistic characteristics to improve job performance rather than addressing barriers within the workplace itself \cite{scott2019factors}, while syntheses of autistic workers' own experiences identify the disabling conditions in environments built around neurotypical colleagues \cite{thorpe2024lived}. Among the terminations those adults attributed to their own work performance, the study reports consistent difficulties rather than declining performance, which an employer could have anticipated or accommodated \cite{pezzimenti2024termination}. The problem therefore sits after hiring and inside the workplace.

Autistic and IDD workers must act on what a workplace leaves unstated. Autistic adults' intolerance of uncertainty is associated with anxiety \cite{hwang2020anxiety}. Employment research documents indirect instructions and feedback that workers may struggle to interpret \cite{mcknightlizotte2018communication}, while accommodation research describes disclosure and adjustment processes that workers find complex to navigate \cite{lindsay2021disclosure,petty2023workplace}. Each study names a separate problem, but one condition runs through all of them, and it is the condition an employer is least likely to see. A sensory environment or an inaccessible commute is concrete enough to be named and contested, whereas an expectation nobody stated leaves no record until a worker has already been judged against it. We use \emph{workplace ambiguity} as a working term for situations where a worker lacks the explicit information needed to interpret what a workplace expects, how it reads their behavior, how it judges their performance, or how it grants support.

Drawing on prior literature, we organize workplace ambiguity into four initial dimensions that provide a framework for examining caregivers' accounts: instructional, social-norm, evaluative, and navigation ambiguity. \textbf{Instructional ambiguity} concerns uncertainty about tasks, priorities, or changing expectations, often stemming from instructions that lack the detail required to translate them into actionable steps \cite{mcknightlizotte2018communication,thorpe2024lived}. \textbf{Social-norm ambiguity} concerns uncertainty about unwritten expectations for interpersonal conduct \cite{prykehobbes2023masking,szechy2024autism}. \textbf{Evaluative ambiguity} concerns uncertainty about how others interpret performance or behavior, especially when feedback is delayed, softened, or insufficient \cite{mcknightlizotte2018communication}. Finally, \textbf{Navigation ambiguity} concerns uncertainty about existing adjustments, request procedures, and support personnel \cite{lindsay2021disclosure,difrancesco2021value}. Section~\ref{subsec:workplace-ambiguity-rw} develops each dimension against its supporting literature.

We examine these dimensions through family caregivers and employment-support professionals who support autistic and IDD workers in work-like settings (referred to collectively as \emph{caregivers}). Employment research treats these caregivers as sustained participants who help autistic and IDD workers secure and keep jobs, coordinate workplace support, and manage employer relationships \cite{voermans2024informal,difrancesco2021value,schwartzman2025family}. An unstated expectation may never register to a worker as an event worth reporting, whereas a caregiver brought in afterward learns what the employer concluded and can help identify the moment the information or support went missing. Because caregivers negotiate with employers directly, they observe both sides of the exchange that produces the ambiguity. In parallel, the tools built for autistic and IDD users make steps explicit and externalize cues a setting would otherwise leave to the user to infer, which is close to what workplace ambiguity requires, but they were developed largely for learning and daily living rather than for work \cite{kagohara2013using,autismspeaksnd,ma2026chatgpt,haroon2025neurobridge}. Little work examines how caregivers and professionals understand ambiguity after job entry, or what technologies they envision for the situations they help workers navigate. Guided by our four initial dimensions, we ask:

\begin{itemize}[leftmargin=*, nosep]
    \item \textbf{RQ1.} \textit{What workplace ambiguities do caregivers observe in the work experiences of autistic and IDD workers?}
    \item \textbf{RQ2.} \textit{What technology supports do caregivers consider useful for addressing these ambiguities?}
\end{itemize}

To understand how this ambiguity operates in practice, we conducted semi-structured interviews ($N=11$) with family caregivers and employment-support professionals (RQ1). Their accounts revealed five themes of workplace ambiguity, including the hidden curriculum of social norms, difficulty evaluating performance expectations, misjudgment of social boundaries, opacity of formal structures, and institutional barriers to accessing support. As our primary contribution, we synthesize these five themes into a conceptual framework of workplace ambiguity comprising three intersecting dimensions: \textit{evaluative} (unclear performance standing), \textit{procedural} (opaque authorized actions), and \textit{relational} (unspoken social interpretations). Across these dimensions, ambiguity arises not just because information is missing, but because expectations typically become explicit only \textit{after} a worker has acted on a guess. To explore solutions, we concluded each interview with a generative-AI co-design activity (RQ2). Participants proposed three forms of technology support: \textit{ambient real-time behavioral redirection} (discreet cues that make hidden boundaries perceptible in the moment), \textit{technology-mediated feedback processing} (tools translating unclear communication into actionable knowledge), and \textit{physiologically informed socio-emotional guidance} (using bodily signals to prompt self-regulation). Viewed through our framework, these supports fundamentally mediate ambiguity by shifting both \textit{when} information reaches a worker and \textit{who} is empowered to interpret it.

Our study makes three core contributions to HCI research on accessibility and the workplace. First, we introduce the conceptual framework of workplace ambiguity to explain why and how post-hiring environments disable autistic and IDD workers. Second, we provide an empirical account of this critical post-hiring period, showing how workers are routinely forced to absorb the costs of unstated expectations. Third, we expose a core technological tension: tools designed to make the workplace legible can inadvertently make the worker hyper-legible to employers through continuous surveillance and forced disclosure. We argue that to foster an ambiguity-aware environment, workplaces must make expectations explicit before they carry consequences, and assistive technologies must deliver actionable information while remaining strictly accountable to the worker, rather than to the employer.

\section{Related Work}
\label{sec:related-work}
Across the three literatures we draw on, workplaces often leave consequential information unstated, while technology rarely addresses this gap or involves the caregivers and job coaches who help workers navigate it.

\subsection{Workplace Barriers Motivating an Ambiguity Lens}
Autistic and IDD individuals represent groups whose workplace experiences reflect differences in social communication, information processing, and support needs~\cite{thorpe2024lived,eisner2025autistic,szechy2024autism}. Employers often associate autism with strengths like reliability and attention to detail, alongside challenges in interpreting expectations and managing change~\cite{thorpe2024lived,bury2021workplace}. Supported employees with intellectual disabilities are less socially included at work than their coworkers, and the study that reports this traces inclusion to limiting expectations and to employer and coworker attitudes~\cite{fillary2006social}. Colleagues and supervisors also hold workers with intellectual disability back through discrimination subtle enough that no single act is contestable~\cite{cavanagh2021subtle}. IDD workers require varying levels of support with tasks and decision-making, depending on how clearly workplaces communicate roles and accommodations~\cite{almalky2020employment,nevala2019effectiveness}.

Prior work often considers autistic and IDD adults together because they share overlapping support needs, exposure to segregated settings, and structural barriers in accessing competitive employment~\cite{clear2022supported,hedley2017employment}. As youth transition from school to work, they often lack clear information on next steps and changing responsibilities. Families report confusion about when school supports end, how to access employment, and who holds responsibility for ongoing coaching~\cite{shattuck2012postsecondary,snellrood2020stakeholder}. In this work, the term ``autistic and IDD workers'' foregrounds both groups' experiences of ambiguity while acknowledging within-group differences.

Workplace challenges for autistic and IDD workers extend beyond unemployment or underemployment. Many who want to work encounter systemic barriers, such as mismatches between their skills and the roles they obtain~\cite{shattuck2012postsecondary,baldwin2014employment}. At work, navigating sensory environments and unclear expectations often strains both autistic workers and their employers~\cite{bury2021workplace,thorpe2024lived,eisner2025autistic}. For workers with intellectual disabilities, a systematic review finds that whether a job lasts depends on the support available from coworkers, supervisors, and job coaches, and that colleagues who underestimate their capacity recur as an obstacle~\cite{ellenkamp2016work}. Qualitative syntheses show autistic workers negotiating their strengths and vulnerabilities through relationships and available workplace supports~\cite{thorpe2024lived}. ``Good fit'' environments provide explicit expectations and accommodations, whereas ``poor fit'' environments feature vague roles and social friction~\cite{thorpe2024lived,eisner2025autistic}. Although these challenges differ in form, many share a common feature: workers are expected to act while important information about expectations, social norms, performance, or support remains unclear.

\subsection{Workplace Ambiguity among Autistic and IDD Workers}
\label{subsec:workplace-ambiguity-rw}
In this study, workplace ambiguity refers to situations in which expectations, interpretations, or available options remain unclear from the worker's perspective, even when they appear obvious to others~\cite{thorpe2024lived,szechy2024autism}. Organizational research treats role ambiguity as a core workplace stressor~\cite{rizzo1970role}, and disability studies describe related uncertainty for autistic and IDD workers as unclear job descriptions, hidden rules, mixed feedback, or opaque support systems~\cite{myles2004hidden,eisner2025autistic,heinze2025workplace}. Drawing on these strands, we synthesize prior work into four initial dimensions of workplace ambiguity, each capturing a recurring pattern: instructional ambiguity in work tasks, social-norm ambiguity in interactions, evaluative ambiguity in feedback, and navigation ambiguity in supports.

\textit{Instructional ambiguity} describes uncertainty about expected duties at work. Prior work reports that autistic workers encounter unclear job expectations and unpredictable duties, describing confusion when expectations around pace or ``flexibility'' remain implied rather than explicitly stated~\cite{kenny2026tell,thorpe2024lived}.

\textit{Social-norm ambiguity} concerns uncertainty about unwritten expectations for interpersonal conduct, such as how to communicate, behave, or respond to coworkers and supervisors in a particular setting. Unlike instructional ambiguity, which concerns what task or action is expected, and evaluative ambiguity, which concerns how performance is judged, social-norm ambiguity concerns how behavior is socially interpreted. Job developers placing youth with IDD report that employers value attitude, dependability, and communication above job-specific technical skill \cite{scheef2019salient}, so a worker can be judged on conduct no one taught them. Research on workplace masking documents the pressure to conform to interpersonal expectations a setting never states \cite{prykehobbes2023masking}, while Milton's double empathy problem holds that workplace misunderstandings often reflect differences in how people of different neurotypes interpret the same interaction rather than a one-sided autistic deficit \cite{milton2012ontological,szechy2024autism}.

\textit{Evaluative ambiguity} concerns uncertainty about how others judge the worker's performance and the resulting consequences. A review notes that workplaces often communicate expectations in ways autistic workers experience as inconsistent, and that tailored supports and respectful working relationships make those expectations easier to read~\cite{heinze2025workplace}. Autistic workers link unpredictable workplaces to stress and burnout~\cite{kenny2026tell,raymaker2023cog}, and perceived workplace climate is associated with their psychological health~\cite{mueller2026employment}. Indirect feedback, polite criticism, and mixed signals can lead to misinterpretation, confusing workers about their actual standing~\cite{mcknightlizotte2018communication}.

\textit{Navigation ambiguity} refers to uncertainty about existing workplace processes, such as accommodations, worker rights, and organizational systems. Autistic workers report that the onus falls on them to identify and request workplace adjustments, a process many struggle with, often without recognizing what accommodations they are entitled to~\cite{davies2022autistic,heinze2025workplace}. Employment guidelines emphasize mapping ``natural supports'' and translating general rules into concrete steps, such as identifying a safe person to answer questions~\cite{ocali2019strategies,ocalindtips,myles2004hidden}. Other forms of ambiguity may also affect workers, but these four dimensions are the ones prior literature supports well enough to carry into an interview protocol.

\subsection{Technology Support for Autistic and IDD Users}
Digital tools long used to support autistic and IDD adults include planners, visual schedules, social stories, and communication aids~\cite{kagohara2013using,autismspeaksnd}. These tools externalize steps that a setting would otherwise leave to the user to infer~\cite{kagohara2013using,karkhaneh2010socialstories}. Although not always framed as responses to ambiguity, they help users manage unclear expectations and transitions.

Researchers have recently explored large language models and generative AI as supports for communication and self-regulation. HCI research examines how autistic adults appropriate general-purpose chatbots for communication and self-regulation~\cite{ma2026chatgpt}, and how AI might bridge cross-neurotype communication by supporting neurotypical perspective-taking~\cite{haroon2025neurobridge}. At the same time, autistic users of general-purpose chatbots report that such tools can validate an interpretation instead of testing it, and can erase a user's own voice through automated masking~\cite{ma2026chatgpt}.

Despite growing interest, much of this work focuses on general learning or daily living contexts rather than workplace settings \cite{kagohara2013using,autismspeaksnd}. Aside from general chatbot use, we have limited understanding of how technology supports workers navigating ambiguity in actual employment environments \cite{ma2026chatgpt,haroon2025neurobridge}. Employment research, however, shows that workplace support is already distributed across multiple stakeholders. Job coaches participate in supporting employees and coordinating with employers \cite{difrancesco2021value}, while families and other informal network members can play sustained roles in pursuing employment, navigating supports, and responding to barriers \cite{voermans2024informal,schwartzman2025family}. IDD employment research has itself interviewed families and caregivers as proxies about a worker's employment, finding that parents help workers negotiate the fit between a person and a job \cite{petnerarrey2016facilitating}, and accessibility research has treated support workers as both proxies for and co-users of technology built for people with intellectual disabilities \cite{balasuriya2022support}. What such supporters hold is a form of knowledge built through proximity: research on parents of people with profound disabilities describes a tacit capacity to read subtle signs, formed through a long shared history of interaction rather than through training \cite{kruithof2024parents}. Prior work positions caregivers and employment-support professionals as participants in the workplace support system itself, making their accounts useful for examining how ambiguous expectations are interpreted, communicated, and acted on across workers, supporters, and employers \cite{difrancesco2021value,voermans2024informal,schwartzman2025family}.

\section{Methods}
\label{sec:methods}
Building on Section~\ref{sec:related-work}, we interviewed 11 caregivers of autistic and IDD workers and included an AI-supported co-design activity grounded in a workplace ambiguity they had described.

\subsection{Data Collection}
\label{subsec:data-collection}
\textbf{Participants.} We interviewed 11 caregivers who support autistic and IDD workers in current or recent employment or training. Families and support professionals are often directly involved in coordinating and providing employment support, including helping workers navigate workplace expectations, employer relationships, and available supports \cite{voermans2024informal,difrancesco2021value,schwartzman2025family}. We recruited these participants because workplace ambiguity is often interpreted and addressed through the support relationships in which they participate. Their perspectives therefore show how supporters recognize ambiguous situations, respond to them, and negotiate with workplaces, while not substituting for workers' own accounts of their experiences. The caregivers included parents, one grandparent caregiver, a job coach, and support professionals. Ten participants reported years of experience between 13 and 30. We recruited through disability service organizations, supported employment providers, and autism-focused community groups in our region via email listservs, social media posts, and direct outreach via organizational staff. A brief screening confirmed that each caregiver currently or recently supported an autistic or IDD person in paid work, volunteer work, or other work-like training. We sought diversity in care relationships (family versus professional) and work settings (e.g., retail, food service, office, sheltered or supported work). Table~\ref{tab:participants} summarizes the participant information.

\begin{table}[t]
  \centering
  \caption{Caregiver and employment-support participant information. Please note that role descriptions are generalized from recruitment records to avoid identifying participants.}
  \label{tab:participants}
  \scriptsize
  \begin{tabularx}{\textwidth}{p{0.01\textwidth}p{0.42\textwidth}p{0.08\textwidth}Y}
    \toprule
    \textbf{ID} & \textbf{Caregiver or support role} & \textbf{Experience} & \textbf{Perspective represented} \\
    \midrule
    \pid{P1} & Job coach and rotating employment-support roles & 22 years & Professional workplace-support perspective \\
    \pid{P2} & Multiple support roles in remote assistive technology and vocational support & 27 years & Professional assistive-technology and vocational-support perspective \\
    \pid{P3} & Remote assistive-technology and vocational-support professional & 21 years & Professional assistive-technology and employment-support perspective \\
    \pid{P4} & Grandparent caregiver & 21 years & Family caregiver perspective \\
    \pid{P5} & Parent caregiver & 22 years & Family caregiver perspective \\
    \pid{P6} & Multiple caregiver and professional support roles & 30 years & Family and professional support perspective \\
    \pid{P7} & Parent caregiver & Not reported & Family caregiver perspective \\
    \pid{P8} & Parent caregiver & 14 years & Family caregiver perspective \\
    \pid{P9} & Parent caregiver & 13 years & Family caregiver perspective \\
    \pid{P10} & Parent caregiver & 27 years & Family caregiver perspective \\
    \pid{P11} & Parent caregiver & 27 years & Family caregiver perspective \\
    \bottomrule
  \end{tabularx}
\end{table}

\textbf{Semi-Structured Interviews with an AI Co-Design Activity.} We conducted 60- to 90-minute interviews via Zoom. Each interview opened with the work history and typical support needs of the worker that caregiver supported. We focused on how caregivers and employment-support professionals recognize, interpret, and respond to workplace ambiguity through ongoing support relationships.
 We then asked caregivers to describe concrete ambiguous workplace episodes, using follow-up probes to clarify context, actions, and perceived consequences. Core prompts targeted instructional, social-norm, evaluative, and navigation ambiguity. We did not present these dimension labels to participants, instead asking caregivers to ``tell a story'' about unclear expectations, social cues, feedback, or support. Table~\ref{tab:protocol} in Appendix~\ref{app:protocol} reports the core questions and probes for each domain.

The co-design activity addressed RQ2. Asking caregivers what technology they wanted in the abstract would have produced a wish list detached from any particular workplace, so we asked each participant to design a support for an episode they had just recounted, which kept every proposed concept tied to a concrete ambiguity. During the same session, caregivers used a large language model (e.g., ChatGPT or Gemini) to explore and refine technology supports for ambiguity they had described. We deployed the AI as a technology probe \cite{hutchinson2003technology} and provisional design material \cite{engstrom2025llm}, in line with speculative design's aim to open debate on possible futures rather than solve present problems \cite{dunneraby2013speculative}. HCI researchers can apply speculative design empirically by engaging participants to react to speculative circumstances and generate knowledge \cite{elsden2017speculative,kozubaev2020expanding}. Grounding this exploration in the caregiver's own ambiguous episode provided a familiar perceptual bridge \cite{auger2013speculative}, building on precedents for conversational AI with disability-adjacent populations \cite{choi2025imagining} and iterative prompt refinement \cite{lc2023speculative}. For one or two situations, caregivers prompted the AI for advice or clarification, reviewed its responses, and refined or critiqued the outputs. The activity revealed how caregivers interpreted ambiguity, preferred support formats, and expected AI to behave at work, including reactions to speculative concepts.

\textbf{Ethics and Data Handling.} Our institution's Institutional Review Board approved the study. Participants provided informed consent before each interview and could decline any question or stop at any time. We audio-recorded and transcribed all interviews using Zoom's built-in automated transcription, then reviewed and corrected transcripts before analysis. We removed or pseudonymized identifying details about caregivers, autistic and IDD workers, employers, and specific workplaces during transcription and analysis. Each participant received a \$40 digital gift card.

\subsection{Data Analysis: Reflexive Thematic Analysis}
\label{subsec:data-analysis}

We applied reflexive thematic analysis (RTA) \cite{braun2006using,braun2019reflecting} primarily inductively to the interview transcripts and AI-supported co-design artifacts, while using the four initial dimensions as a limited deductive lens when examining ambiguity-related data. RTA is a method for developing and interpreting patterns of meaning in qualitative data that treats researchers' prior knowledge and social positions as a resource for interpretation rather than as bias to be removed \cite{braun2019reflecting}. We selected RTA because our goal was to characterize recurring forms of workplace ambiguity across caregivers' accounts, and because recognizing an episode as ambiguous, rather than as a performance failure, required familiarity with disability employment support. Two coders coded the data. The first read all transcripts and assigned initial codes to every relevant excerpt, and the second, senior to the first, met with them weekly to review the codes one by one against their original quotations, so that coding decisions were examined as they accumulated rather than audited at the end. The two coders grouped the resulting codes into sub-themes on an affinity diagram and then consolidated sub-themes into primary themes. This process produced 153 initial codes, 13 sub-themes, and five primary themes for RQ1, and 54 initial codes, 10 sub-themes, and three primary themes for RQ2, all reported in Appendix~\ref{app:codebook}. For example, the codes ``rules change with each supervisor'' and ``policies intentionally written vague'' were grouped into the subtheme \emph{Context-Dependent Unwritten Social Norms}, which we report in Section~\ref{subsubsec:the-hidden-curriculum-of-workplace-social-norms}, while codes such as ``uncertain when conversation has gone on too long'' and ``missing nonverbal redirect cues'' formed \emph{Difficulty Calibrating Social Engagement with the Public} in Section~\ref{subsubsec:misjudgement-of-social-boundaries-with-non-disabled-peers}. Disagreements concerned where a code belonged rather than what it meant. The coders initially split closely related codes about misread humor across two themes, placing ``sarcasm misread as genuine anger'' under performance expectations and ``difficulty calibrating response to humor and sarcasm'' under social boundaries. The coders resolved this disagreement by returning to the quotation and asking which uncertainty the worker faced at that moment, then keeping the code with the theme that named that uncertainty.

Because RQ2 asks what technology could address the ambiguity documented in RQ1, we coded the two questions against each other. Our interview protocol asked caregivers to propose supports for a workplace episode they had just described, so each technology idea was grounded in a specific situation. When coding an excerpt for RQ2, we recorded which RQ1 data points the technology addressed, following the participant's own account of the problem rather than our judgment of what such a technology could do in principle. We then revisited each idea to ask whether it also addressed ambiguities that the same participant had raised elsewhere in the interview; on this basis, 25 of the 54 RQ2 codes were linked to more than one ambiguity. Aggregating these links to the theme level yields the correspondence reported in Table~\ref{tab:rq1-rq2-map}. Details regarding researcher positionality can be found in Appendix~\ref{sec:researcher-positionality}.

\section{Findings}
\label{sec:findings}
Section~\ref{subsec:rq1-ambiguity} reports the five themes of workplace ambiguity caregivers observed, and Section~\ref{subsec:rq2-technology} reports the three technology-support themes participants proposed and shows how they relate to the ambiguity themes. Table~\ref{tab:rq1-rq2-map} shows which ambiguity themes each technology support addressed. We use \emph{dimensions} for the analytic categories we carry into and out of the study, and \emph{themes} for what the analysis produced from participants' perspectives.

\subsection{RQ1: Themes of Workplace Ambiguity}
\label{subsec:rq1-ambiguity}

Ambiguity at work is not unique to autistic and IDD workers. The expectations that mattered were conveyed by inference rather than by statement, inferring wrongly brought a disciplinary consequence instead of a moment of awkwardness, and asking for clarification carried a social cost that made checking risky in itself. Where those conditions held together, information a workplace withheld operated as an access barrier, not an ordinary workplace annoyance.

\subsubsection{The Hidden Curriculum of Workplace Social Norms}
\label{subsubsec:the-hidden-curriculum-of-workplace-social-norms}

The hidden curriculum captures unwritten relational knowledge workers must navigate alongside their formal jobs. Caregivers and colleagues rarely taught this curriculum directly, yet colleagues still used it to evaluate autistic and IDD workers. Ambiguity arose when workers had to infer which informal rule mattered and how others would interpret their behavior. As \pid{P1} described, one worker experienced employment as two simultaneous jobs:

\begin{quote}
One of the most difficult aspects of his job was that he felt like he had to drive in two lanes at the same time, one of those being the technical aspects of his job, and one of them being the social aspects of his job. And for him, he could do this, or he could do that. But he can't do both at the same time.
\end{quote}

\pid{P1}'s description shows that the assigned task was not the only site of difficulty: the worker also had to track relational standing while doing the work. Although his direct supervisor gave positive feedback, friction with another workplace relationship contributed to a performance improvement plan. Ambiguity was therefore relationally distributed: doing the assigned job well could still leave a worker vulnerable if they did not know which secondary relationships mattered. The hidden curriculum also shifted with local workplace culture. As \pid{P3} explained:

\begin{quote}
He never had problems with breaks. He knows it's 30 minutes. Things that are really black and white are great. It's those unwritten social norms you pick up by just living.
\end{quote}

\pid{P3}'s contrast between black-and-white break rules and ``unwritten social norms'' shows how workplace culture asks workers to infer expectations that are never formally taught. Participants described this inference work across everyday interactions: \pid{P3} described a worker copying truck drivers' language and being reprimanded, while \pid{P4} described a worker completing a delivery task correctly but missing the small exchange that made the interaction feel polite.

Literal compliance created a further double bind. Workers could be praised for following rules during onboarding but penalized when strict rule-following became socially inconvenient. \pid{P4}'s account of a worker repeatedly reporting ``blocked exits'' and ``moldy products'' shows how formal safety rules could become risky when local workplace culture expected those rules to be softened. The hidden curriculum was a workplace access problem in which autistic and IDD workers had to infer when formal rules should be softened by informal expectations.

\subsubsection{Difficulty in Evaluating Performance Expectations}
\label{subsubsec:difficulty-in-evaluating-performance-expectations}

The difficulty in evaluating performance expectations describes the uncertainty workers faced when judging whether their work was sufficient, correct, or responsive to changing expectations. Participants described supervisors assuming that unstated standards, changes, or feedback cues would be inferred, while the workers they supported were relying on what had actually been said. Performance became a moving target when instructions, quality standards, feedback timing, and help-seeking norms were communicated indirectly. \pid{P11} described how far an instruction could fall short of what a worker needed:

\begin{quote}
If you say just make a pizza, it's just a blank stare. I don't know, just some of the things that we would at least start with, they might have a challenge with because they don't want to do it incorrectly.
\end{quote}

The instruction can appear clear while leaving the worker without procedural anchors. The worker may need the sequence, stopping point, and permission to ask follow-up questions. \pid{P11} made the same point about multi-step directions, noting that ``one thing is good'' but two or three steps may require workers to return and ask again.

The act of help-seeking also operated as an ambiguous social behavior. Participants described workers hesitating to seek clarification where asking could bring a reprimand or make them appear incompetent, and so attempting the task perfectly instead. Feedback therefore mattered, but participants said it was often delayed or softened, losing its practical impact. As \pid{P5} emphasized:

\begin{quote}
It's of absolutely no use for a person with autism to wait and come back later and give feedback after they've finished something, or give feedback on something they did three weeks ago.
\end{quote}

Timing mattered because workers could not easily connect delayed comments to the behavior that needed adjustment. \pid{P1} similarly noted that feedback could sound like ``a nice-to-have'' rather than something that would ``make or break my job.'' When supervisors avoided direct correction, workers lost the chance to learn the actual performance standard before formal consequences appeared.

Performance ambiguity was both temporal and social. Expectations varied across trainers, task shifts, and workplace rhythms, while signals of poor performance often arrived too late. Participants described how a lack of time awareness, visible overwhelm, and emotional overcorrection could interact with unclear workplace expectations and contribute to misunderstandings, including instances of managers interpreting confusion as an attitude. Workplace expectations often became legible only after workers had already been judged to have failed.

\subsubsection{Misjudgment of Social Boundaries with Non-Disabled Peers}
\label{subsubsec:misjudgement-of-social-boundaries-with-non-disabled-peers}

Misjudgment of social boundaries with non-disabled peers centers on uncertainty about where professional interaction should begin, stop, or change across people and settings. Participants described autistic and IDD workers as facing particular difficulty when workplaces conveyed these boundaries through subtle social cues rather than explicit guidance. Participants often framed this difficulty as a problem of invisible cues rather than a lack of desire to relate. As \pid{P3} explained:

\begin{quote}
Reading people's body language doesn't come easy for a lot of people on the spectrum, especially, but generally, people with IDD. Boundaries are invisible. You can't see them, but you have to figure that out.
\end{quote}

These boundaries, such as when to end an interaction or how much discretion a worker has, do exist, but workplaces often leave them to be inferred through social cues. Non-disabled coworkers and customers may use subtle embodied cues to signal that a conversation should end, while autistic and IDD workers may miss those cues or interpret them too literally. \pid{P2} noted that a ``get-back-on-task'' nod may be obvious to some workers but require explanation for others.

Boundary miscalibration was often handled through discipline rather than mentoring. A worker might trust a coworker's permission to use a company car, only to receive an official warning because that coworker lacked the necessary authority. Another might repeat a joke after the first laugh because laughter seemed like encouragement, not a signal to stop. Relationship-building was also part of workplace participation. \pid{P11} noted that some task-driven workers approach work merely as work, completing the job without small talk or friendship-building. Their coworkers may then misread this attitude as disinterest. The boundary problem also involved knowing when to engage enough to be read as collegial.

\subsubsection{Opacity of Formal Workplace Structures}
\label{subsubsec:opacity-of-formal-workplace-structures}

The opacity of formal workplace structures refers to the difficulty of reading institutional procedures that appear standardized but are actually inaccessible in practice. For autistic and IDD workers in participants' accounts, this became an accessibility barrier when navigating these structures depended on independently inferring unstated roles, procedures, or sources of authority. Workers were expected to navigate formal systems independently, even when they did not know who had authority to answer a question, where the relevant process lived, or what language made a request legitimate.

Participants emphasized that formal roles were often split across different people without clear explanation. As \pid{P1} described: ``There might be multiple roles in a single place. This person is gonna be the one who trains you on the functions of your job, but this other person is the one you call off to, and this other person is the one you talk to about benefits.'' \pid{P1}'s example shows that workplace structure can be ambiguous even when every role has a formal owner. Knowing the trainer does not mean knowing who handles absences, benefits, HR questions, or policy exceptions. \pid{P11} described a similar problem with self-service systems for requesting time off: ``nobody shows you that kind of stuff,'' leaving workers ``kind of left on your own.'' The barrier was not motivation, but the expectation that workers could translate scattered procedures into action.

The stakes of procedural opacity were high. \pid{P4} described a worker who asked about using bereavement time to visit a dying relative. When the relative had not yet died, and the worker came to work, management interpreted the request as dishonesty and fired him.

Accommodation processes created similar ambiguity. Workers were often not taught how to mention accommodations, request them, or transfer them across management changes. \pid{P1} observed that schools may produce ``beautiful documents'' listing accommodations without teaching workers how to ask for comparable support at work. Formal structures became barriers when they assumed procedural literacy: policies existed, but pathways to them were difficult to see.

\subsubsection{Barriers to Accessing Workplace Support}
\label{subsubsec:barriers-to-accessing-workplace-support}
Support barriers emerged when a recognized need still had to get past practical, social, and institutional obstacles. Workers and caregivers often had to negotiate through several gatekeepers before an employer would act. The ambiguity was practical and political: workers needed to determine whether support was allowed, whether it would be resented, and whether it would remain available when circumstances changed. Participants described institutional pushback as a recurring barrier. As \pid{P1} recounted:

\begin{quote}
She came to me, and I put her in contact with human resources, who immediately rejected her meeting. She's requesting an accommodation. We are legally required. And HR got my boss involved, and then she yelled at me and told me I should stop encouraging my employees.
\end{quote}

This account shows that access to support could function as gatekeeping rather than as a simple request for help. Access to support depended on whether HR and supervisors understood disability law and recognized support requests as legitimate. In another case, \pid{P1} described a situation where a supervisor removed a job coach from the email recipient list and refused further contact. Such responses depleted workers' and advocates' energy, sometimes pushing workers to leave rather than continue fighting for support.

Even when support was allowed, it could make workers stand out in ways that led to stigmatization. \pid{P3} described a situation where coworkers asked why a worker received ``all this extra help,'' requiring others to explain that the support was not an advantage but a way to access less natural work expectations. Noticeable job coaching, sensory tools, or accommodations could cause peers to judge the worker through non-disabled standards. Support could therefore resolve one ambiguity while creating another: the question of how to receive help without the stigma of being deemed less competent.

Support could also become either too thin or too intrusive. Fragmented job coaching often failed to cover the moments when ambiguity occurred during an actual shift. Families absorbed the labor of customized support when employers did not provide usable onboarding training. Yet \pid{P6} cautioned that overcompensation could prevent workers from developing confidence in uncertain situations: ``Because everyone is overcompensating all the time, a lot of times people jump in too fast. That person did not even have the opportunity to feel the uncertainty. They don't figure it out because everyone's doing too much too quick.'' \pid{P6}'s concern clarifies that effective support must be timely, legitimate, and autonomy-preserving. Workplace ambiguity persists when support systems are reactive, socially stigmatized, or dependent on individual advocates rather than built into workplace routines.

\subsection{RQ2: Technology Supports for Workplace Ambiguity}
\label{subsec:rq2-technology}

Our analysis of the co-design data produced three technology-support themes: real-time cueing that makes hidden expectations visible in the moment, feedback-processing tools that translate workplace communication and procedures into actionable knowledge, and socio-emotional guidance that helps workers and supporters notice patterns of stress or escalation. No technology-support theme corresponded to a single ambiguity theme, and 25 of the 54 RQ2 codes were linked to more than one ambiguity (Section~\ref{subsec:data-analysis}). They were proposing changes to when workplace information arrives and who helps interpret it, and a change of that kind reaches several ambiguities at once. Table~\ref{tab:rq1-rq2-map} reports which ambiguity themes each support addressed. Real-time cueing and feedback processing each reached all five themes, but they reached them differently. Real-time cueing concentrated on hidden social norms, performance expectations, and social boundaries, and touched the two institutional themes only once each. Feedback processing accumulated the most links of any support and was the only one participants proposed by more than a single caregiver in every column, because they treated unclear communication as a barrier that recurs across domains. Physiologically informed support reached only performance expectations and social boundaries. Bodily signals may help a worker anticipate overwhelm, but they cannot say which procedure applies or who controls access to support.

\begin{table}[t]
  \centering
  \caption{Which workplace ambiguity themes each technology support addressed. Rows are the three RQ2 technology-support themes; columns are the five RQ1 ambiguity themes. Each cell lists the participants whose proposed support addressed that ambiguity, following each participant's own account of the problem the support would solve. The final column reports how many of the five ambiguity themes each support reached and how many participant--ambiguity links it accumulated.}
  \label{tab:rq1-rq2-map}
  \scriptsize
  \setlength{\tabcolsep}{3.5pt}
  \begin{tabularx}{\textwidth}{@{}>{\raggedright\arraybackslash}p{0.135\textwidth}YYYYY>{\raggedright\arraybackslash}p{0.088\textwidth}@{}}
    \toprule
    \textbf{RQ2 technology support}
    & \multicolumn{5}{c}{\textbf{RQ1 workplace ambiguity theme}}
    & \textbf{Reach} \\
    \cmidrule(lr){2-6}
    & \textbf{Hidden curriculum} (\ref{subsubsec:the-hidden-curriculum-of-workplace-social-norms})
    & \textbf{Performance expectations} (\ref{subsubsec:difficulty-in-evaluating-performance-expectations})
    & \textbf{Social boundaries} (\ref{subsubsec:misjudgement-of-social-boundaries-with-non-disabled-peers})
    & \textbf{Formal structures} (\ref{subsubsec:opacity-of-formal-workplace-structures})
    & \textbf{Accessing support} (\ref{subsubsec:barriers-to-accessing-workplace-support})
    & \\
    \midrule

    \textbf{Ambient real-time behavioral redirection} (\ref{subsubsec:ambient-real-time-behavioral-redirection})
    & P4, P10
    & P1, P3, P4, P5, P6, P10
    & P2, P3, P4
    & P4
    & P1
    & 5 of 5\newline 13 links \\
    \addlinespace[3pt]

    \textbf{Technology-mediated feedback processing} (\ref{subsubsec:technology-mediated-feedback-processing})
    & P4, P7, P11
    & P1, P4, P5, P6, P8, P10, P11
    & P3, P5, P7, P9, P11
    & P4, P6, P11
    & P1, P5
    & 5 of 5\newline 20 links \\
    \addlinespace[3pt]

    \textbf{Physiologically informed socio-emotional guidance} (\ref{subsubsec:physiologically-informed-socio-emotional-guidance})
    & ---
    & P1, P2, P3, P4, P5, P6
    & P3
    & ---
    & ---
    & 2 of 5\newline 7 links \\

    \bottomrule
  \end{tabularx}
\end{table}

\subsubsection{Ambient Real-Time Behavioral Redirection}
\label{subsubsec:ambient-real-time-behavioral-redirection}

Ambient real-time behavioral redirection describes discreet technologies that make hidden workplace boundaries perceptible at the moment they matter. These tools could cue timing, proximity, sensory load, or task transitions without requiring a job coach to intervene publicly. Their promise was immediacy: many ambiguities only become visible when a worker is already interacting with a customer, moving through a task, or trying to regulate attention.

Representative artifacts included conversation-duration wearables, proximity detectors, sensory-environment controls, and task-goal prompting systems. Across them, technology made otherwise invisible limits visible before a supervisor or customer reacted negatively.

\begin{figure}[!htbp]
  \centering
  \includegraphics[width=0.6\textwidth]{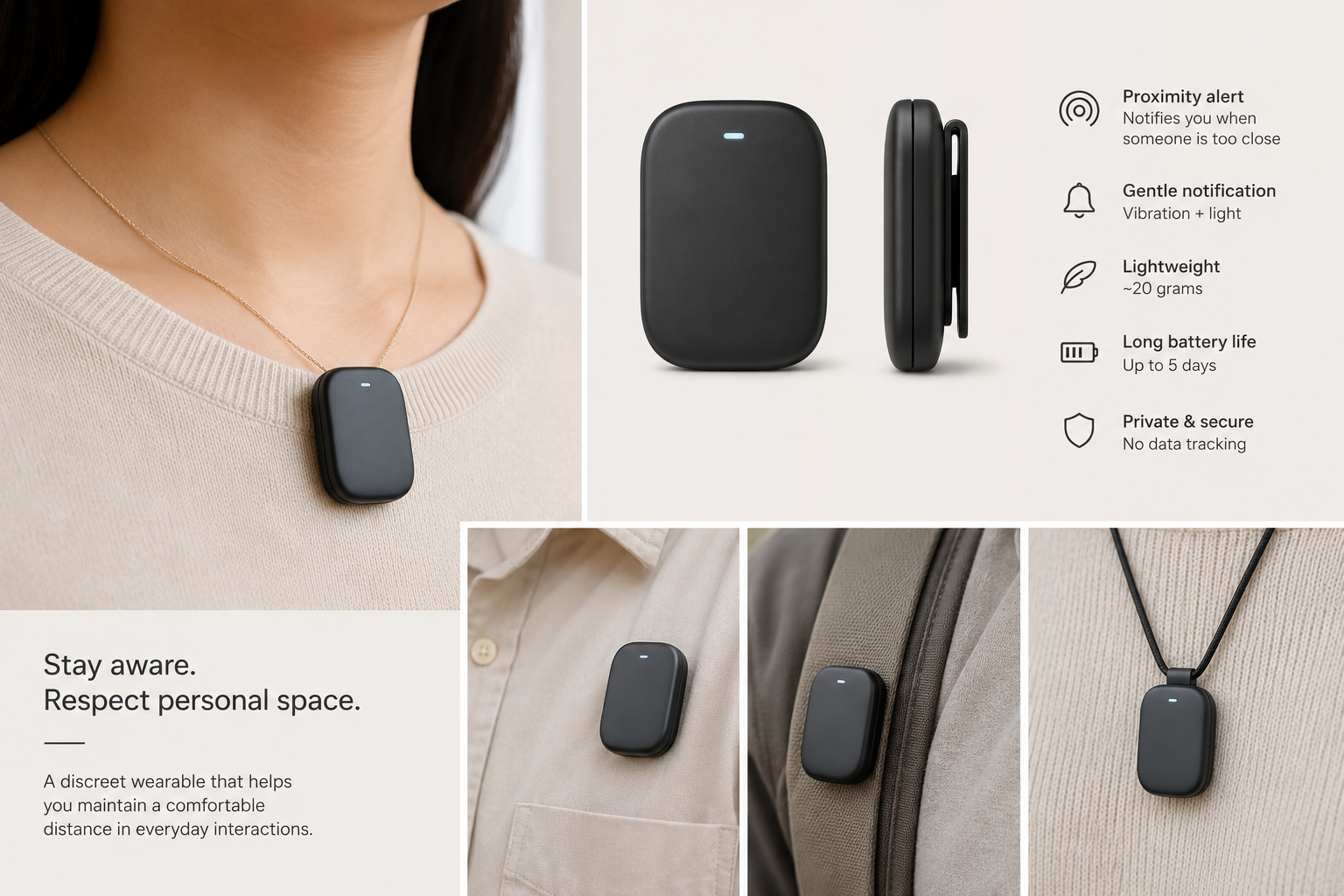}
  \caption{P3 artifact showing a discreet wearable proximity cue, iteratively refined into a smaller brooch-like form intended to provide subtle feedback when interpersonal distance becomes too close.}
  \Description{AI-supported co-design artifact depicting a small circular wearable attached to clothing. The device is designed to provide a subtle proximity alert when another person is too close.}
  \label{fig:proximity-wearable}
\end{figure}

The iterative design process also surfaced discretion as part of accessibility rather than a secondary aesthetic concern. \pid{P3} refined their \emph{proximity wearable} concept from a more noticeable device into a smaller, brooch-like form intended to blend into workplace clothing (Figure~\ref{fig:proximity-wearable}). This refinement illustrates a tension across real-time cueing concepts: making an invisible workplace boundary perceptible to the worker without making the support itself conspicuous to others.

\begin{figure}[!htbp]
  \centering
  \includegraphics[width=0.6\textwidth]{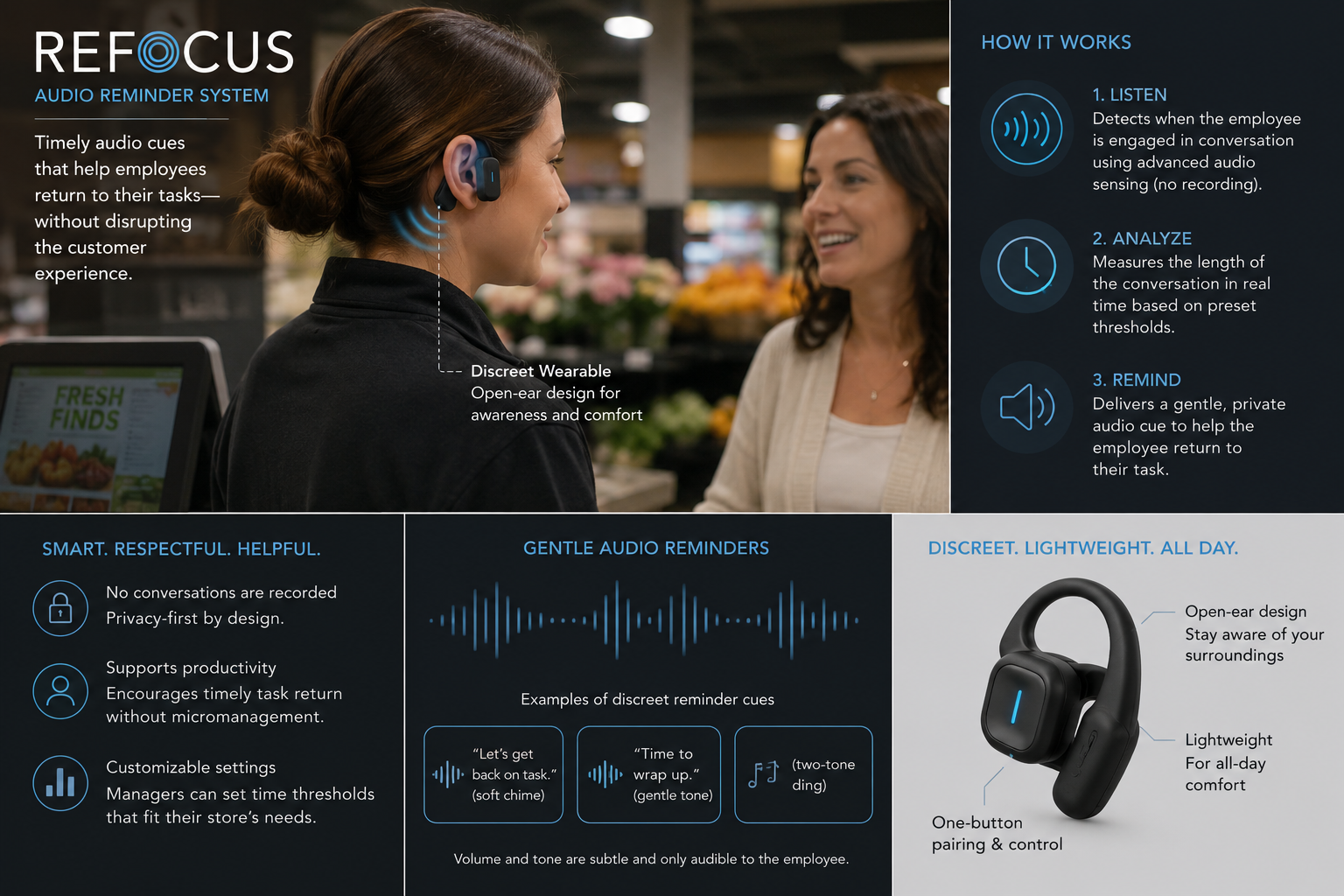}
  \caption{P2 artifact showing Refocus, a discreet audio reminder system intended to help a worker return to task after extended customer interaction.}
  \Description{Artifact screenshot of Refocus, a discreet audio reminder system for returning to task after extended customer interaction.}
  \label{fig:refocus}
\end{figure}

Participants most often described these tools as gentle cueing systems rather than corrective surveillance. \pid{P2}'s \emph{Refocus} concept, for example, used discreet audio reminders to help a worker return to task after extended customer interactions (Figure~\ref{fig:refocus}). As \pid{P2} explained:

\begin{quote}
I would want it to be able to gently, kindly remind them to stay on task, probably with audio cues. If a customer goes through checkout in 3 minutes average, your conversation probably shouldn't be 3 minutes or something.
\end{quote}

\pid{P2}'s example shows how technology could translate an implicit social boundary into an actionable cue. The worker would not have to infer from a customer's body language that the conversation had gone too long. A discreet reminder could make the expected rhythm of the task visible, with similar supports for proximity, breaks, task switching, and getting unstuck from repetitive work.

Real-time redirection also made temporal expectations concrete. \pid{P4} described a system that distinguished between ``6:52 is good, 7:03 is good, but 7:07 is too late.'' Such specificity matters because vague time expectations often become disciplinary only after the worker has crossed an invisible line. Other concepts used sensory-environment monitoring to help workers notice noise, heat, lighting, or crowding before overload disrupted work.

Cueing was not an easy fix. Alerts had to be individualized, sensory-safe, and discreet enough not to stigmatize the worker. Repeated alarms could be turned off, while overly literal prompts could lead to overcorrection. \pid{P3} worried that a worker might take a cue so literally that they would ``never do it again.'' Some production environments also ban personal devices.

\subsubsection{Technology-Mediated Feedback Processing}
\label{subsubsec:technology-mediated-feedback-processing}

The theme of technology-mediated feedback processing captures tools that translate ambiguous workplace communication into accessible, actionable knowledge. Unlike the previous theme of real-time redirection, this theme focused on interpretation after information had already been communicated unclearly. It also included employer-facing tools, shifting responsibility from worker adaptation toward a more accessible communicative environment.

The artifacts included feedback translators, employer-facing communication guides, HR or policy question interfaces, and tools for identifying work interests. Rather than proposing more reminders, these concepts reorganized how workplace knowledge is explained and routed.

\begin{figure}[!htbp]
  \centering
  \includegraphics[width=0.6\textwidth]{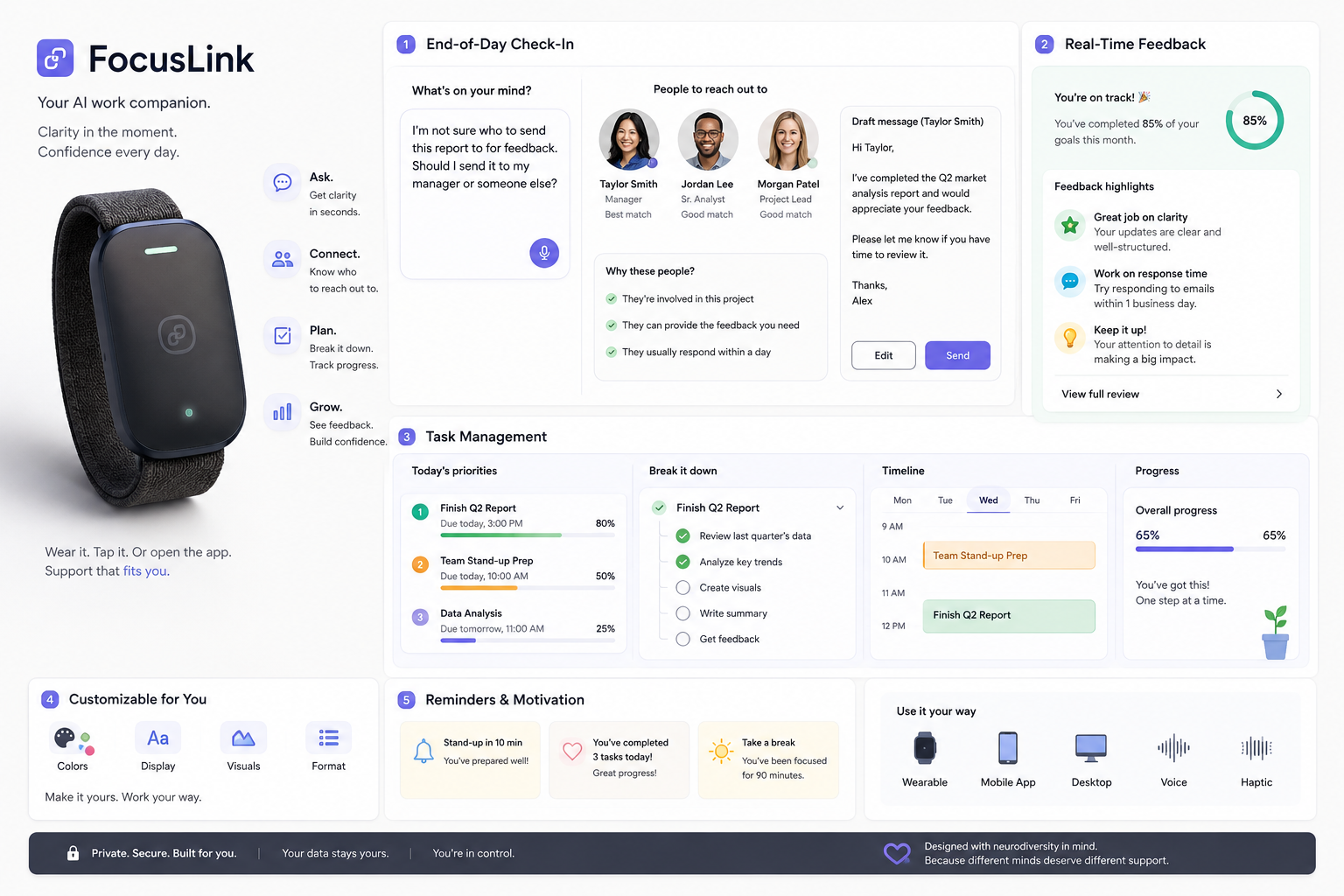}

  \caption{P6 artifact showing FocusLink, a work companion concept that combines question routing, real-time feedback, task management, and confidence support.}
  \Description{Artifact screenshot of FocusLink, a work companion concept with question routing, real-time feedback, task management, and confidence supports.}
  \label{fig:focuslink}
\end{figure}

Feedback translation was a core use case. One representative artifact was \pid{P6}'s \emph{FocusLink}, a work companion concept combining question routing, real-time feedback, task management, and confidence support (Figure~\ref{fig:focuslink}). \pid{P1} similarly said such a tool could help with ``Taking feedback from supervision,'' applying it, and reporting back. Workers may hear feedback without knowing whether it is urgent, how it changes the next task, or whether follow-up is expected. Automated check-ins and dashboards could make performance status visible before discipline occurs, giving workers and supervisors a shared view of unresolved issues. Participants also saw value in plain-language policy support. As \pid{P11} described:

\begin{quote}
Wouldn't it be great if they could ask questions back? Because, you know, we're all embarrassed about asking questions. But boy, if it was here, you know, like HR-loaded questions, boy, that would probably benefit them too.
\end{quote}

\pid{P11}'s suggestion turns on the social risk of asking for help. A question interface could reduce embarrassment by allowing workers to explore HR or policy issues before approaching someone. \pid{P11} also imagined businesses loading job-specific procedures into a system so a worker could select a task, read the steps, and demonstrate understanding. It was a low-risk way to clarify the terms of work. Participants further argued that technology should help employers communicate in a more accessible way. As \pid{P4} proposed:

\begin{quote}
On the employer side, if they had something that said, okay, my employee has autism. If I say this, this, and this, what does that mean to him? How might they perceive what I'm telling them to do?
\end{quote}

This shifts the design target from fixing the worker to repairing communication between worker and workplace. Tools could help managers phrase instructions more clearly and route questions to the right person. Yet social coaching fit training settings better than live work. \pid{P9} said a tool could be helpful in the classroom because ``nobody will feel weird,'' but useless in real interaction if it distracts the worker from the person speaking. Technology-mediated feedback processing thus has to balance interpretation, privacy, and stigma while making workplace knowledge more legible.

\subsubsection{Physiologically Informed Socio-Emotional Guidance}
\label{subsubsec:physiologically-informed-socio-emotional-guidance}

Physiologically informed socio-emotional guidance centers on technologies that use bodily or contextual signals to help workers notice emotional escalation before it becomes a workplace crisis. Internal state awareness mattered because workers may not know when their body is signaling the need for a break, a different strategy, or help from a supervisor.

\begin{figure}[!htbp]
  \centering
  \includegraphics[width=0.6\textwidth]{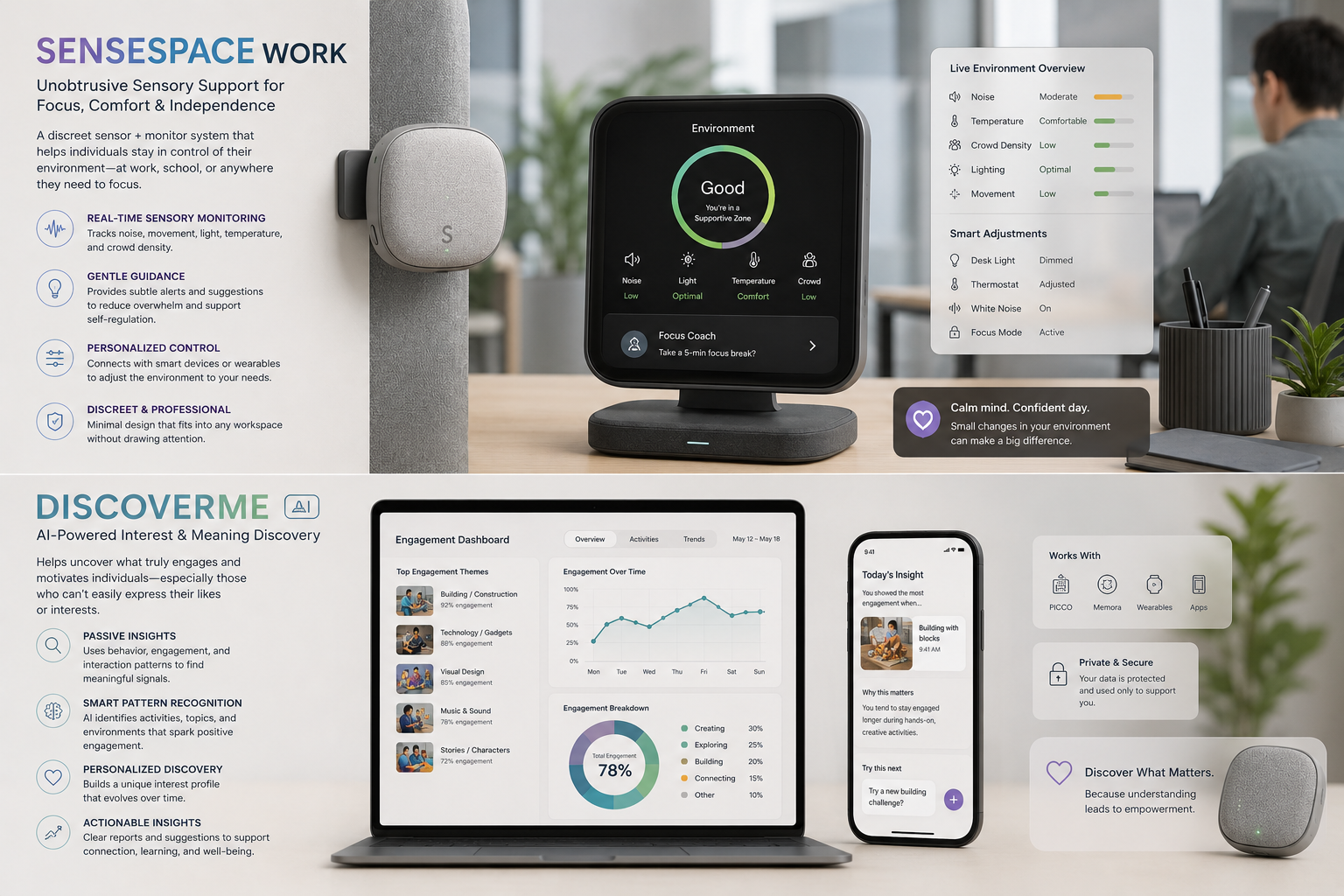}
  \caption{P5 artifact showing SenseSpace Work, an unobtrusive sensor and monitor concept for tracking sensory conditions and prompting regulation.}
  \Description{Artifact screenshot of SenseSpace Work, an unobtrusive sensor and monitor concept for tracking sensory conditions and prompting regulation.}
  \label{fig:sensespace}
\end{figure}

Representative artifacts used biofeedback prompts, self-regulation cues, and confidence supports during task transitions. They framed emotion and bodily state as part of the work environment rather than as private issues separate from job performance. Participants imagined biofeedback as a way to support self-regulation before difficult interactions. As \pid{P1} described:

\begin{quote}
Notice your heartbeat is higher than usual. You might need to take a break before you talk to your supervisor. That would be super awesome. You might need to go self-regulate because it feels like you're a little discombobulated right now.
\end{quote}

The system would not label the worker as anxious. It would help the worker decide when to pause before entering a high-stakes conversation. Similar transition supports, such as a ``pep talk buddy,'' could remind a worker that they had been trained, could ask for help, and had handled similar situations before.

Physiological data could also help supporters understand patterns after the workday. \pid{P6} described using data to identify when anxiety spikes happen, such as during lunch, breaks, or specific task transitions. \pid{P5}'s \emph{SenseSpace Work} concept imagined unobtrusive monitoring of sensory conditions to support regulation before distress escalated (Figure~\ref{fig:sensespace}), rather than leaving caregivers to intervene only after harm or meltdown. Technology could therefore make emotional patterns discussable without relying only on memory or crisis reports. However, participants repeatedly warned that physiological data are themselves ambiguous. As \pid{P5} asked:

\begin{quote}
I don't know how the technology can discern excitement versus anxiety if they're both increasing heart rate and blood flow to your skin or organs or whatever.
\end{quote}

\pid{P5}'s concern limits what designers can claim biofeedback knows. The same signal may mean anxiety, excitement, effort, or environmental stress depending on the person and moment. Participants therefore called for individualized thresholds and worker control over who can see the data. Without those limits, physiological guidance could become workplace surveillance rather than support. Bodily data may help workers and supporters identify patterns, but only if the system treats interpretation as uncertain, personal, and accountable to the worker.

\section{Discussion}
\label{sec:discussion}

Across the five themes, workers had to act before needed workplace information became legible. What remained unclear varied across episodes. In some, a worker could not tell how their own behavior would be read; in others, they could not tell how they were doing or what a supervisor's feedback required of them; in still others, they could not tell what to do next or whose answer counted. The three RQ2 technology themes responded by changing when information reached workers and who could help interpret it (Sections~\ref{subsubsec:ambient-real-time-behavioral-redirection}--\ref{subsubsec:physiologically-informed-socio-emotional-guidance}). 

Prior work documents these barriers separately. Employment research describes indirect communication and unclear job expectations for autistic workers \cite{thorpe2024lived,petty2023workplace,mcknightlizotte2018communication}, and employment-support research documents how employees, employers, and job coaches sustain support after hiring \cite{difrancesco2021value}. Rizzo and colleagues, working in organizational psychology, name a related condition as role ambiguity, where employees lack clear information about their job responsibilities, authority, or performance expectations \cite{rizzo1970role}. What connects these strands is a question none asks directly: \textit{when does missing information become a barrier to access rather than simply an inconvenience?} We develop that question through three refined dimensions of workplace ambiguity in Section~\ref{subsec:multidimensional-access}, and then examine how technology mediates the work of autistic and IDD employees with their workplaces in Section~\ref{subsec:technology-mediation}. We derive design and practice guidelines for an ambiguity-aware workplace from our participants' insights in Section~\ref{subsec:design-practice-guidelines}.

\subsection{Workplace Ambiguity as a Multi-Dimensional Access Problem}
\label{subsec:multidimensional-access}

\begin{figure}[t]
  \centering
  \includegraphics[width=0.7\textwidth]{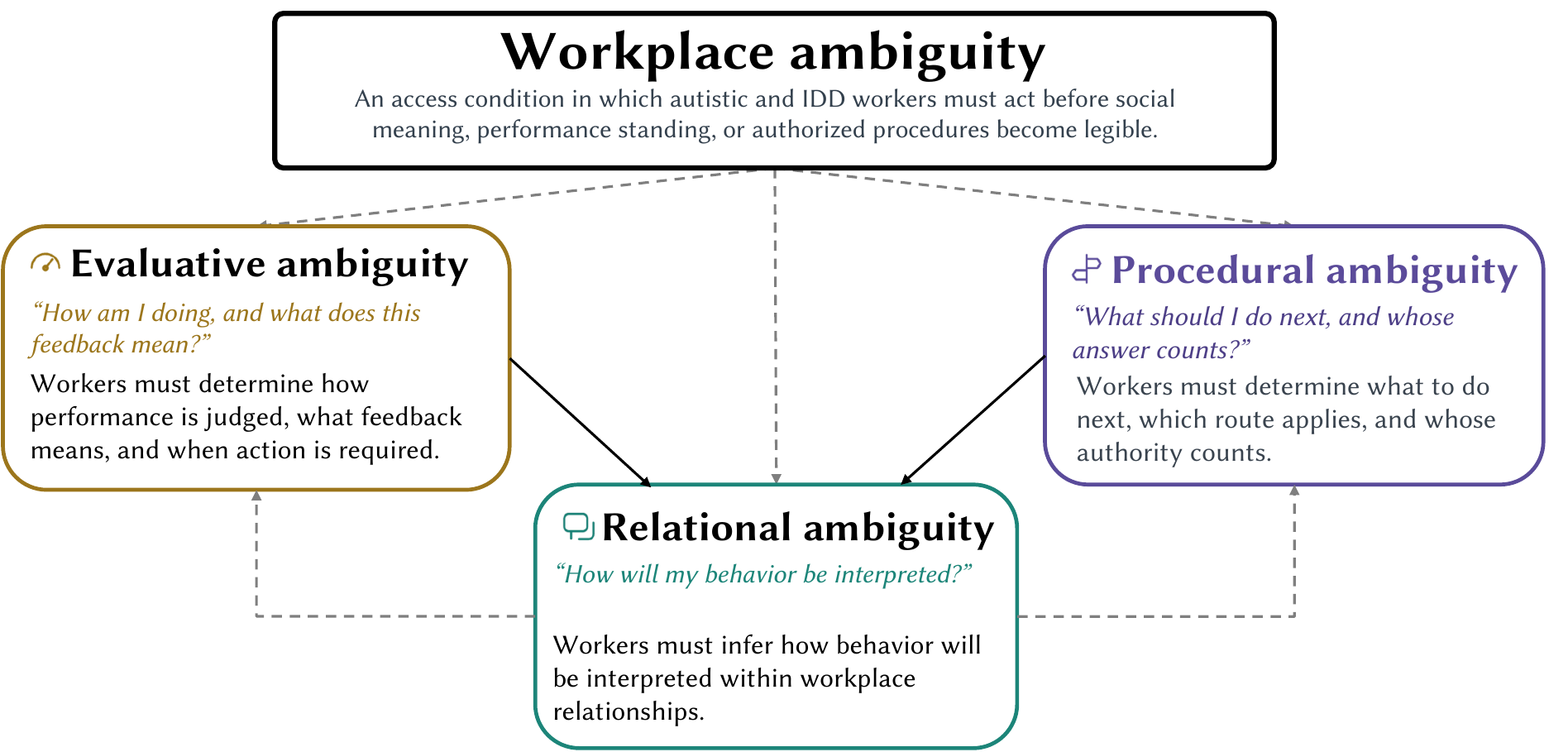}
  \caption{Workplace ambiguity as a multi-dimensional access problem. Evaluative and procedural ambiguity can become relationally consequential, while relational ambiguity can in turn reinforce evaluative and procedural ambiguity.}
  \Description{A diagram defining workplace ambiguity as an access condition in which workers must act before consequential information becomes legible. Workplace ambiguity includes evaluative ambiguity about performance standing and feedback, procedural ambiguity about next steps and authority, and relational ambiguity about how behavior will be interpreted. Arrows from evaluative and procedural ambiguity point toward relational ambiguity, while arrows from relational ambiguity return toward evaluative and procedural ambiguity, showing how these dimensions can reinforce one another.}
  \label{fig:ambiguity-dimensions}
\end{figure}

We began with four initial dimensions drawn from prior literature: instructional, social-norm, evaluative, and navigation ambiguity (Section~\ref{subsec:workplace-ambiguity-rw}). The five themes we identified through interviews with caregivers in Section~\ref{subsec:rq1-ambiguity} did not map onto these dimensions one-to-one. (1) Findings about performance stayed closest to evaluative ambiguity but sharpened it by focusing on whether usable feedback arrived while a worker could still act. (2) Findings about formal structures and access to support combined instructional and navigation concerns around a higher-order question of which next action the workplace would recognize as authorized, producing procedural ambiguity. (3) Findings about hidden curricula and social boundaries centered on how workplace actors interpret behavior, producing relational ambiguity. Instructional ambiguity did not survive as a dimension of its own. It divided, with uncertainty about what counts as correct performance joining evaluative ambiguity and uncertainty about which action an employer will authorize joining procedural ambiguity. Together, these findings refine the four initial dimensions into evaluative, procedural, and relational ambiguity. The novelty is not in noticing that work can be unclear, since organizational psychology named role ambiguity more than fifty years ago \cite{rizzo1970role}. Role ambiguity describes a stressor in how a job is designed, measured across a general workforce, and asks what unclear expectations do to an employee's wellbeing. Our dimensions ask who bears the consequence when a workplace leaves consequential information unstated, and they treat the three conditions as inseparable, because what an employer fails to make evaluatively or procedurally clear gets settled socially by whoever interprets the worker in the meantime. Although these dimensions of ambiguity can affect any worker, our findings show how they become accessibility concerns for autistic and IDD workers when workplaces rely on implicit expectations, indirect feedback, or social cues that workers are expected to infer. This framing also extends employment research beyond whether autistic and IDD workers obtain a job \cite{shattuck2012postsecondary,baldwin2014employment,hedley2017employment} to what happens after a job begins. Our findings show that continued workplace access is shaped by whether workers can recognize when their job standing is at risk, how colleagues and supervisors interpret their behavior, and how caregivers become involved when workplace expectations or procedures are unclear. Employment access is not achieved once at hiring; it is continually shaped through workplace interactions. Figure~\ref{fig:ambiguity-dimensions} summarizes this three-dimensional framework.

\textbf{Evaluative ambiguity.} First, evaluative ambiguity concerns whether autistic and IDD workers can determine where they stand and what a given piece of feedback requires them to do. Participants described instructions that appeared complete to supervisors but left workers uncertain about what counted as correct or sufficient performance (Section~\ref{subsubsec:difficulty-in-evaluating-performance-expectations}). Feedback delivered after a task could also arrive too late for workers to connect it to the behavior requiring adjustment, while softened feedback could obscure whether a message was routine coaching or a consequential warning. Organizational research identifies unclear performance standards as a structural source of role ambiguity \cite{rizzo1970role}, while employment research documents how indirect instruction and unpredictable expectations affect autistic employees \cite{mcknightlizotte2018communication,heinze2025workplace,kenny2026tell}. We extend this work by showing that evaluative access depends on timing as well as clarity. Feedback can be clear yet inaccessible when it arrives too late for workers to respond. The clearest instance in our findings occurred when a formal consequence became the first unambiguous signal that performance had been at risk.

\textbf{Procedural ambiguity.} Second, procedural ambiguity concerns which next action an employer will recognize as authorized and whose approval makes it so. Participants described responsibilities divided across separate offices, so that the person who trained an autistic or IDD worker was not the person they contacted when absent, and neither of them handled benefits (Section~\ref{subsubsec:opacity-of-formal-workplace-structures}). They also described supervisors and human resources staff blocking or delegitimizing support requests, which left caregivers to reconstruct the applicable policy before a request could be made (Section~\ref{subsubsec:barriers-to-accessing-workplace-support}). These findings bring instructional and navigation ambiguity together around what an employer will accept as the correct move. Prior work on workplace adjustments shows that autistic workers carry the burden of identifying and requesting them \cite{davies2022autistic,lindsay2021disclosure,heinze2025workplace}, while employment-support research documents how caregivers and job coaches coordinate access to that support \cite{difrancesco2021value,voermans2024informal,schwartzman2025family}. We show that formally available support remains inaccessible when workers cannot safely activate it.

\textbf{Relational ambiguity.} Third, relational ambiguity concerns how coworkers, supervisors, and clients will interpret an autistic or IDD worker's behavior. Participants described workers managing their technical responsibilities while tracking a second, largely unstated social job (Section~\ref{subsubsec:the-hidden-curriculum-of-workplace-social-norms}). They also described local expectations and ``invisible'' boundaries that became consequential only once someone read a worker's behavior differently than the worker intended, as shown in Section~\ref{subsubsec:misjudgement-of-social-boundaries-with-non-disabled-peers}. Relational ambiguity therefore broadens the social-norm ambiguity in our initial framework beyond uncertainty about unwritten rules. The problem is that others interpret the behavior, often after the fact, and those interpretations carry organizational weight. Prior work on the hidden curriculum describes social rules that a setting assumes rather than teaches \cite{myles2001understanding,myles2004hidden}, while Milton's double empathy problem holds that autistic and non-autistic people may interpret the same interaction differently \cite{milton2012ontological,szechy2024autism}. Our findings identify that problem within employment, where an unstated expectation can shape judgments of a worker's competence even after that worker has completed the formal task \cite{bury2021workplace,prykehobbes2023masking}.

Relational ambiguity is also where evaluative and procedural ambiguity can become socially consequential. When feedback arrives too late to act on, a manager may read the resulting confusion or overcorrection as attitude rather than uncertainty (Section~\ref{subsubsec:difficulty-in-evaluating-performance-expectations}). When a procedure is unclear, a supervisor may read a worker's attempt to navigate it as dishonesty or noncompliance (Section~\ref{subsubsec:opacity-of-formal-workplace-structures}). Across these accounts, evaluative and procedural ambiguity became consequential through the social interpretations attached to workers' behavior. The relationship also runs in the other direction. Ambiguity about how questions will be received may discourage autistic and IDD workers from seeking clarification, leaving performance expectations opaque, a pattern consistent with research linking intolerance of uncertainty to anxiety among autistic adults \cite{hwang2020anxiety,raymaker2023cog}. Similarly, when coworkers treat visible support as evidence of lesser competence, workers may avoid support routes the employer formally offers (Section~\ref{subsubsec:barriers-to-accessing-workplace-support}). Relational ambiguity can emerge from and reinforce evaluative and procedural ambiguity, which matters because perceived workplace climate is closely tied to autistic employees' psychological health \cite{mueller2026employment}.

This interdependence changes where intervention must act. For workers, clarifying a procedure or performance standard does not necessarily resolve what colleagues concluded about them while the information was unclear. For colleagues and supervisors, an interpretation formed during a moment of ambiguity can become part of a worker's standing and shape later access to feedback or support. Conceptually, the three dimensions connect literatures that have described these problems separately. Employment research documents unclear expectations \cite{thorpe2024lived,petty2023workplace} and the burden of securing adjustments \cite{davies2022autistic,heinze2025workplace}. Organizational psychology conceptualizes role ambiguity in terms of job expectations and authority \cite{rizzo1970role}. Milton's account of double empathy explains cross-neurotype misreading \cite{milton2012ontological,szechy2024autism}, without asking what an employer subsequently does with it. Framing workplace ambiguity as an access problem brings these accounts together. Evaluative and procedural ambiguity concern information not yet made usable; relational ambiguity concerns what others conclude meanwhile.

\textbf{Treating ambiguity as an access problem moves the burden from a worker's inference toward organizational practices that support workers' continued, equal participation in employment.} It would be a weaker claim to say that workplaces should become unambiguous. The design goal is therefore not to eliminate ambiguity altogether, but to make consequential ambiguity recognizable and actionable before it produces harm. Much of what participants described as unwritten is also what makes a workplace a social setting rather than a script, and they were not asking for every expectation to be codified. What workers lacked, in these accounts, was not a complete rulebook but a way to find out in time which unstated expectation was about to matter. Our view of that gap is necessarily secondhand because it comes through caregivers' accounts. We can see the gap because caregivers were already interpreting on workers' behalf, so their interpretive labor is simultaneously our evidence and part of the arrangement we describe. Reading workplace ambiguity through caregivers therefore reveals how employers may distribute the work of interpretation without acknowledging it, while leaving open how the same conditions appear to the workers who experience them. Design research has long treated ambiguity as a resource that opens interpretation rather than a defect to remove \cite{gaver2003ambiguity}, and anti-solutionist work resists framing every condition as a problem awaiting a fix \cite{blythe2016antisolutionist}. We meet that tradition from accessibility, where the ambiguity at stake is not one a designer introduces but one an employer leaves in place, and where its cost falls on a particular worker.  

\subsection{Technology Support as Mediation Between Autistic and IDD Workers and Workplaces}
\label{subsec:technology-mediation}

Every concept participants proposed changed when workplace information reached a worker, who else could help interpret it, or both (Section~\ref{subsec:rq2-technology}). Each therefore mediates between an autistic or IDD worker and their workplace. Verbeek's technological mediation framework treats technologies as shaping perception and action rather than operating as neutral channels \cite{verbeek2005what}. Assistive-technology research demonstrates the value of externalized cues and structured support \cite{kagohara2013using,klavina2024assistive,fage2016tablet}, while collaborative employment research shows that support for a worker already extends to their family members and job coaches \cite{ara2024collaborative,difrancesco2021value}. None of these concepts has been built, so we treat them as proposals for changing access to workplace information after job entry rather than as validated systems \cite{hutchinson2003technology,elsden2017speculative}.

Real-time cueing makes a hidden boundary perceptible before a worker is penalized for missing it. Participants proposed cues tied to how long a conversation had run or how close a worker was standing, each offering discreet guidance while a situation was still unfolding, as shown in Section~\ref{subsubsec:ambient-real-time-behavioral-redirection}. These concepts extend research on structured task support into employment settings, where cueing occurs within relationships involving performance judgment and unequal authority \cite{kagohara2013using,klavina2024assistive}. Timing is what makes them valuable, since a cue can make an implicit limit actionable while a worker can still clarify, adjust, or self-regulate. The same property creates risk. A visible cue can disclose disability status, a literal cue can encourage overcorrection, and continuous cueing can become behavioral monitoring. Real-time mediation therefore expands access only when workers control it and when the behavior it nudges toward is not simply neurotypical behavior \cite{milton2022ten}. In our findings, this form of support clustered most strongly around evaluative and relational ambiguity because the same cue that makes a task threshold visible can also make an otherwise implicit social boundary perceptible before it becomes consequential.

Feedback processing distributes clarification among the people already involved instead of leaving workers to resolve ambiguity alone. Participants proposed feedback translators and employer-facing communication guides, alongside policy interfaces a worker could consult before approaching a person (Section~\ref{subsubsec:technology-mediated-feedback-processing}). These concepts build on collaborative employment research by treating clarification as a shared workplace process rather than only a worker-facing aid \cite{ara2024collaborative,difrancesco2021value,voermans2024informal}. A tool that only helps workers translate themselves into existing norms leaves the communication that produced the ambiguity unchanged. Bidirectional clarification could instead help employers communicate expectations more concretely, allow caregivers to contribute relevant context, and give workers lower-risk ways to ask questions. However, LLM-based tools can reframe language without knowing the authority, policy, or consequence attached to a message, which is why a job coach in prior work questioned the advice such a tool produced \cite{jang2024llm}. Feedback processing must therefore leave workers in control of what is disclosed and of who is present when it is. Participants proposed feedback processing across more ambiguity links than any other support (Table~\ref{tab:rq1-rq2-map}), because it can clarify performance expectations, surface procedural routes and authority, and reduce the one-sided inference behind relational ambiguity.

Socio-emotional guidance supports reflection while preserving boundaries around interpretation. Participants imagined physiological prompts that could encourage workers to pause before high-stakes interactions, while warning that the same signal might reflect anxiety or excitement (Section~\ref{subsubsec:physiologically-informed-socio-emotional-guidance}). Physiological data can indicate arousal without establishing its cause, just as AI advice can sound authoritative without accounting for local policy, relationship history, or organizational authority. These limitations establish a boundary of interpretive humility. Technologies may help workers notice patterns, consider explanations, or identify someone who can clarify a situation, but they should not determine another person's intent or classify a worker's emotional state. Research on LLM support and cross-neurotype communication identifies both the value and risk of interpretive assistance \cite{jang2024llm,ma2026chatgpt,haroon2025neurobridge}. These tools are most consistent with ambiguity-aware support when they preserve uncertainty and remain accountable to autistic and IDD workers. In our data, socio-emotional guidance concentrated on evaluative ambiguity and reached relational ambiguity only once, because physiological signals may help workers recognize overload before performance or interaction consequences emerge, yet they cannot determine which organizational procedure applies or who has authority to approve a request.

Each of the three themes carries a risk that follows from the same property that makes it useful. The central question is therefore not simply whether technology reduces workplace ambiguity, but how it redistributes information, responsibility, and power for autistic and IDD workers. In our co-design activity, the LLM also functioned as a mediating design material that prompted caregivers to articulate, evaluate, and refine how support might operate in the workplace. Caregivers can help contextualize technologies and assess whether they fit a worker's needs, but their involvement should inform rather than replace the worker's authority. Workers should retain control over what a technology does on their behalf and over who may read what it produces. Employers carry an obligation of their own because a support tool that can be repurposed for discipline or productivity monitoring stops functioning as support.

\subsection{Design and Practice Guidelines for Ambiguity-Aware Workplaces}
\label{subsec:design-practice-guidelines}

Three of the four guidelines below address evaluative, procedural, and relational access in turn. The fourth addresses the risk that runs across all three, namely that a technology built to make the workplace legible can instead make the worker legible to the workplace.

\textbf{Make evaluation usable when it is given.} Feedback should reach autistic and IDD workers while they can still act on it, and it should distinguish routine guidance from consequential evaluation. Participants described instructions that stopped short of what a worker needed and feedback that arrived too late for that worker to connect it to the relevant behavior (Section~\ref{subsubsec:difficulty-in-evaluating-performance-expectations}). Supervisors should state the applicable standard, the required adjustment, and whether the message constitutes coaching or a formal concern. Feedback systems can preserve this context and create a shared record that workers may review with a chosen supporter. Plain-language translation is insufficient if it strips a message of its urgency or its authority. AI tools may organize or explain feedback, but only the workplace can confirm a supervisor's intent or a formal consequence. Timing is part of accessibility because information that arrives after the actionable moment may be clear in content but inaccessible in practice.

\textbf{Make routes and authority traceable.} Workers and caregivers should be able to see authorized next steps and who has the authority to approve them. The procedural findings show workers and caregivers navigating fragmented roles and challenging gatekeepers who blocked help the employer formally offered (Sections~\ref{subsubsec:opacity-of-formal-workplace-structures} and~\ref{subsubsec:barriers-to-accessing-workplace-support}). An employer's systems should identify which procedure applies and who can authorize a decision, and should say what a worker can do when the answer is no. Workers should not need specialized terminology, advanced knowledge of the correct office, or unnecessary disclosure to begin seeking support \cite{lindsay2021disclosure}. Caregivers may help when invited, but employers remain responsible for keeping procedures accurate and usable. Shared process representations can reduce advocate labor by making organizational routes visible rather than requiring caregivers to reconstruct them informally.

\textbf{State the expectations you evaluate against.} Workplaces should state the locally consequential social expectations they actually evaluate against without using technology to enforce neurotypical behavioral conformity. The relational findings show that coworkers and supervisors evaluated autistic and IDD workers through unstated expectations even when workers completed formal tasks successfully (Sections~\ref{subsubsec:the-hidden-curriculum-of-workplace-social-norms} and~\ref{subsubsec:misjudgement-of-social-boundaries-with-non-disabled-peers}). Supervisors should explain local expectations before evaluating a worker against them, and coworkers should communicate boundaries directly rather than relying only on subtle cues. Workplaces should also provide private ways to request clarification when expectations change. Caregivers can help identify recurring ambiguity when workers invite their participation, but they should not define acceptable behavior on the worker's behalf. The hidden curriculum and the double empathy problem together explain why explicitness should reduce one-sided inference without normalizing behavior \cite{myles2001understanding,myles2004hidden,milton2012ontological,szechy2024autism}.

\textbf{Keep interpretation accountable to the autistic and IDD worker.} Ambiguity-aware technologies should clarify consequential workplace information without turning autistic and IDD workers into objects of continuous interpretation. Participants valued discreet, individualized support while raising concerns that clustered on a single axis, namely who would see the data and what would be concluded from it (Sections~\ref{subsubsec:ambient-real-time-behavioral-redirection}--\ref{subsubsec:physiologically-informed-socio-emotional-guidance}). Workers should control when assistance is active, what information a system collects, and who can access it. Caregivers and employment-support professionals may help configure a system when invited, but they should not receive data by default or override worker choices. Employers should prohibit the use of support data for productivity scoring, discipline, or emotion classification. AI systems should separate workplace source material from generated interpretation and communicate uncertainty when context is incomplete \cite{ma2026chatgpt}. Technology supports workplace legibility only when autistic and IDD workers retain meaningful authority over how it interprets and represents them.

Our findings shift the design problem from helping autistic and IDD workers become better at inferring inaccessible workplaces toward making consequential workplace information available before workers bear the consequences of not knowing it. Evaluative ambiguity concerns performance standing and the consequences embedded in feedback, procedural ambiguity concerns the action route and source of authority, and relational ambiguity concerns how coworkers and supervisors interpret a worker's behavior. Technology can mediate each condition, but it can also reproduce inaccessible workplaces when it increases surveillance, imposes conformity, or substitutes technological inference for legitimate communication and organizational responsibility. An ambiguity-aware workplace therefore does not ask workers to become progressively better at inferring what an organization leaves unsaid; it makes socially, evaluatively, and procedurally consequential information accessible in time to act on it.

\section{Limitations and Future Work}
Our study has limitations that inform future work. First, our investigation of workplace ambiguity is secondhand. Proxy accounts are known to diverge from self-report in this population. Facility staff rate the quality of life of adults with intellectual disabilities lower than those adults rate it themselves, and agreement between the two is poor \cite{koch2015proxy}, while agreement across reporters is weakest on interpersonal relationships and self-determination \cite{berastegui2021reporters}. Caregivers described what they observed and what they did in response, so we can report the situations workers encountered and the consequences that followed, but not how those situations felt from the inside or what a worker concluded in the moment (Section~\ref{subsec:data-collection}). Future work can examine the three dimensions directly with autistic and IDD workers and ask whether the interdependence we describe in Section~\ref{subsec:multidimensional-access} appears the same from workers' own perspectives. Second, the technology concepts we report are speculation rather than evidence. Participants produced them in a single session using a generative AI tool, so what they imagined was shaped by what the model returned, and no concept here has been built or evaluated \cite{hutchinson2003technology,elsden2017speculative}. We also heard one side of the workplace. Supervisors, human resources staff, and coworkers appear in our data as the people who participants described as withholding information or blocking a request, never as actors explaining why an organization leaves what it leaves unsaid. Studying that side would test whether workplace ambiguity is neglect, deliberate discretion, or an artifact of how work is divided across offices. Building even one of these supports would then raise the harder question our participants kept returning to, which is not whether a system can surface workplace information but who is permitted to act on what it surfaces.

\section{Conclusion}
We interviewed 11 caregivers of autistic and IDD workers and asked them to design technology supports for the workplace situations they had just described. Their accounts identify five themes of workplace ambiguity and three technology-support themes. We synthesize the five ambiguity themes into evaluative, procedural, and relational ambiguity and read these dimensions as a problem of access rather than of individual inference. We show that these dimensions feed one another, that employment access is renegotiated in everyday interaction rather than settled at hiring, and that a technology built to clarify a workplace can as easily be turned on the worker. A workplace that expects a worker to guess has already decided what it owes them, and naming that decision is what makes it arguable.

\bibliographystyle{ACM-Reference-Format}
\bibliography{main}

@article{lord2020autism,
  author    = {Lord, Catherine and Brugha, Traolach S. and Charman, Tony and Cusack, James and Dumas, Guillaume and Frazier, Thomas and Jones, Emily J. H. and Jones, Rebecca M. and Pickles, Andrew and State, Matthew W. and Taylor, Julie Lounds and Veenstra-VanderWeele, Jeremy},
  title     = {Autism Spectrum Disorder},
  journal   = {Nature Reviews Disease Primers},
  year      = {2020},
  volume    = {6},
  number    = {1},
  pages     = {5},
  doi       = {10.1038/s41572-019-0138-4},
  url       = {https://doi.org/10.1038/s41572-019-0138-4}
}

@article{schalock2021overview,
  author  = {Schalock, Robert L. and Luckasson, Ruth and Tass{\'e}, Marc J.},
  title   = {An Overview of {Intellectual Disability: Definition, Diagnosis, Classification, and Systems of Supports} (12th ed.)},
  journal = {American Journal on Intellectual and Developmental Disabilities},
  year    = {2021},
  volume  = {126},
  number  = {6},
  pages   = {439--442},
  doi     = {10.1352/1944-7558-126.6.439},
  url     = {https://doi.org/10.1352/1944-7558-126.6.439}
}

@article{petty2023workplace,
  author  = {Petty, Stephanie and Tunstall, Lydia and Richardson, Hannah and Eccles, Niamh},
  title   = {Workplace Adjustments for Autistic Employees: What Is ``Reasonable''?},
  journal = {Journal of Autism and Developmental Disorders},
  year    = {2023},
  volume  = {53},
  number  = {1},
  pages   = {236--244},
  doi     = {10.1007/s10803-021-05413-x},
  url     = {https://doi.org/10.1007/s10803-021-05413-x}
}

@article{difrancesco2021value,
  author  = {Di Francesco, Cynthia and Murahara, Flavio and Martin, Valerie and Flanagan, Tara and Nadig, Aparna},
  title   = {The Value of Employment Support Services for Adults on the Autism Spectrum and/or with Intellectual Disabilities: Employee, Employer, and Job Coach Perspectives},
  journal = {Journal of Vocational Rehabilitation},
  year    = {2021},
  volume  = {55},
  number  = {3},
  pages   = {283--296},
  doi     = {10.3233/JVR-211163},
  url     = {https://doi.org/10.3233/JVR-211163}
}

@article{hwang2020anxiety,
  author  = {Hwang, Ye In (Jane) and Arnold, Samuel and Srasuebkul, Preeyaporn and Trollor, Julian},
  title   = {Understanding Anxiety in Adults on the Autism Spectrum: An Investigation of Its Relationship with Intolerance of Uncertainty, Sensory Sensitivities and Repetitive Behaviours},
  journal = {Autism},
  year    = {2020},
  volume  = {24},
  number  = {2},
  pages   = {411--422},
  doi     = {10.1177/1362361319868907},
  url     = {https://doi.org/10.1177/1362361319868907}
}

@article{mcknightlizotte2018communication,
  author  = {McKnight-Lizotte, Michelle},
  title   = {Work-Related Communication Barriers for Individuals with Autism: A Pilot Qualitative Study},
  journal = {The Australian Journal of Rehabilitation Counselling},
  year    = {2018},
  volume  = {24},
  number  = {1},
  pages   = {12--26},
  doi     = {10.1017/jrc.2018.4},
  url     = {https://doi.org/10.1017/jrc.2018.4}
}

@article{lindsay2021disclosure,
  author  = {Lindsay, Sally and Osten, Victoria and Rezai, Mana and Bui, Sunny},
  title   = {Disclosure and Workplace Accommodations for People with Autism: A Systematic Review},
  journal = {Disability and Rehabilitation},
  year    = {2021},
  volume  = {43},
  number  = {5},
  pages   = {597--610},
  doi     = {10.1080/09638288.2019.1635658},
  url     = {https://doi.org/10.1080/09638288.2019.1635658}
}

@article{prykehobbes2023masking,
  author  = {Pryke-Hobbes, Amber and Davies, Jade and Heasman, Brett and Livesey, Adam and Walker, Amy and Pellicano, Elizabeth and Remington, Anna},
  title   = {The Workplace Masking Experiences of Autistic, Non-Autistic Neurodivergent and Neurotypical Adults in the {UK}},
  journal = {PLOS ONE},
  year    = {2023},
  volume  = {18},
  number  = {9},
  pages   = {e0290001},
  doi     = {10.1371/journal.pone.0290001},
  url     = {https://doi.org/10.1371/journal.pone.0290001}
}

@article{voermans2024informal,
  author  = {Voermans, Moniek A. C. and Wilthagen, Ton and Embregts, Petri J. C. M.},
  title   = {Informal Network Members' Perspectives and Experiences on Work for People with Intellectual Disabilities: A Thematic Synthesis},
  journal = {Journal of Occupational Rehabilitation},
  year    = {2024},
  volume  = {34},
  number  = {1},
  pages   = {37--55},
  doi     = {10.1007/s10926-023-10128-0},
  url     = {https://doi.org/10.1007/s10926-023-10128-0}
}

@article{schwartzman2025family,
  author  = {Schwartzman, Benjamin C. and Lanchak, Emily R. and Carter, Erik W. and Guest, Laura E. and McMillan, Elise D. and Taylor, Julie L.},
  title   = {Family Perspectives on Promoting Paid Employment for Individuals With Intellectual and Developmental Disabilities},
  journal = {Rehabilitation Counseling Bulletin},
  year    = {2025},
  volume  = {68},
  number  = {2},
  pages   = {79--91},
  doi     = {10.1177/00343552231213996},
  url     = {https://doi.org/10.1177/00343552231213996}
}

@article{klavina2024assistive,
  author  = {Klavina, Aija and P{\'e}rez-Fuster, Patricia and Daems, Jo and Lyhne, Cecilie N{\o}rby and Dervishi, Eglantina and Pajalic, Zada and {\O}derud, Tone and Fuglerud, Kristin S. and Markovska-Simoska, Silvana and Przybyla, Tomasz and Klichowski, Michal and Stiglic, Gregor and Laganovska, Egija and Alarc{\~a}o, Soraia M. and Tkaczyk, Alan H. and Sousa, Carla},
  title   = {The Use of Assistive Technology to Promote Practical Skills in Persons with Autism Spectrum Disorder and Intellectual Disabilities: A Systematic Review},
  journal = {DIGITAL HEALTH},
  year    = {2024},
  volume  = {10},
  pages   = {20552076241281260},
  doi     = {10.1177/20552076241281260},
  url     = {https://doi.org/10.1177/20552076241281260}
}

@article{fage2016tablet,
  author    = {Fage, Charles and Pommereau, L{\'e}onard and Consel, Charles and Balland, Emilie and Sauz{\'e}on, H{\'e}l{\`e}ne},
  title     = {Tablet-Based Activity Schedule in Mainstream Environment for Children with Autism and Children with {ID}},
  journal   = {ACM Transactions on Accessible Computing},
  year      = {2016},
  volume    = {8},
  number    = {3},
  articleno = {9},
  numpages  = {26},
  publisher = {Association for Computing Machinery},
  address   = {New York, NY, USA},
  doi       = {10.1145/2854156},
  url       = {https://doi.org/10.1145/2854156}
}

@article{karkhaneh2010socialstories,
  author  = {Karkhaneh, Mohammad and Clark, Brenda and Ospina, Maria B. and Seida, Jennifer C. and Smith, Veronica and Hartling, Lisa},
  title   = {{Social Stories}\texttrademark{} to Improve Social Skills in Children with Autism Spectrum Disorder: A Systematic Review},
  journal = {Autism},
  year    = {2010},
  volume  = {14},
  number  = {6},
  pages   = {641--662},
  doi     = {10.1177/1362361310373057},
  url     = {https://doi.org/10.1177/1362361310373057}
}

@inproceedings{ara2024collaborative,
  author    = {Ara, Zinat and Ganguly, Amrita and Peppard, Donna and Chung, Dongjun and Vucetic, Slobodan and Genaro Motti, Vivian and Hong, Sungsoo Ray},
  title     = {Collaborative Job Seeking for People with Autism: Challenges and Design Opportunities},
  booktitle = {Proceedings of the CHI Conference on Human Factors in Computing Systems},
  series    = {CHI '24},
  year      = {2024},
  articleno = {73},
  numpages  = {17},
  pages     = {1--17},
  location  = {Honolulu, HI, USA},
  publisher = {Association for Computing Machinery},
  address   = {New York, NY, USA},
  isbn      = {9798400703300},
  doi       = {10.1145/3613904.3642197},
  url       = {https://doi.org/10.1145/3613904.3642197}
}

@inproceedings{jang2024llm,
  author    = {Jang, JiWoong and Moharana, Sanika and Carrington, Patrick and Begel, Andrew},
  title     = {{``It's the Only Thing I Can Trust'': Envisioning Large Language Model Use by Autistic Workers for Communication Assistance}},
  booktitle = {Proceedings of the CHI Conference on Human Factors in Computing Systems},
  series    = {CHI '24},
  year      = {2024},
  articleno = {77},
  numpages  = {18},
  pages     = {1--18},
  location  = {Honolulu, HI, USA},
  publisher = {Association for Computing Machinery},
  address   = {New York, NY, USA},
  isbn      = {9798400703300},
  doi       = {10.1145/3613904.3642894},
  url       = {https://doi.org/10.1145/3613904.3642894}
}

@article{almalky2020employment,
  author  = {Almalky, Hussain A.},
  title   = {Employment Outcomes for Individuals with Intellectual and Developmental Disabilities: A Literature Review},
  journal = {Children and Youth Services Review},
  volume  = {109},
  pages   = {104656},
  year    = {2020},
  doi     = {10.1016/j.childyouth.2019.104656},
  url     = {https://doi.org/10.1016/j.childyouth.2019.104656}
}

@misc{autismspeaksnd,
  author       = {{Autism Speaks}},
  title        = {How Digital Tools Can Help},
  howpublished = {Autism and the Transition to Adulthood Tool Kit excerpt},
  year         = {n.d.},
  url          = {https://www.autismspeaks.org/tool-kit-excerpt/how-technology-can-help}
}

@article{baldwin2014employment,
  author  = {Baldwin, Susanna and Costley, Debra and Warren, Anthony},
  title   = {Employment Activities and Experiences of Adults with High-Functioning Autism and {Asperger's} Disorder},
  journal = {Journal of Autism and Developmental Disorders},
  volume  = {44},
  number  = {10},
  pages   = {2440--2449},
  year    = {2014},
  doi     = {10.1007/s10803-014-2112-z},
  url     = {https://doi.org/10.1007/s10803-014-2112-z}
}

@article{bury2021workplace,
  author  = {Bury, Simon M. and Flower, Rebecca L. and Zulla, Rosslynn and Nicholas, David B. and Hedley, Darren},
  title   = {Workplace Social Challenges Experienced by Employees on the Autism Spectrum: An International Exploratory Study Examining Employee and Supervisor Perspectives},
  journal = {Journal of Autism and Developmental Disorders},
  volume  = {51},
  number  = {5},
  pages   = {1614--1627},
  year    = {2021},
  doi     = {10.1007/s10803-020-04662-6},
  url     = {https://doi.org/10.1007/s10803-020-04662-6}
}

@misc{clear2022supported,
  author       = {{Clearinghouse for Labor Evaluation and Research (CLEAR), U.S. Department of Labor}},
  title        = {Supported Employment for People with Intellectual Disabilities, Autism Spectrum Disorders, and Schizophrenia: A Propensity Matched Comparison of Vocational Rehabilitation Outcomes ({Chan} \& {Kregel}, 2019)},
  year         = {2022},
  url          = {https://clear.dol.gov/study-database},
  note         = {Study profile, last reviewed August 2022. Retrieved August 12, 2026}
}

@article{davies2022autistic,
  author  = {Davies, Jade and Heasman, Brett and Livesey, Adam and Walker, Amy and Pellicano, Elizabeth and Remington, Anna},
  title   = {Autistic Adults' Views and Experiences of Requesting and Receiving Workplace Adjustments in the {UK}},
  journal = {PLOS ONE},
  volume  = {17},
  number  = {8},
  pages   = {e0272420},
  year    = {2022},
  doi     = {10.1371/journal.pone.0272420},
  url     = {https://doi.org/10.1371/journal.pone.0272420}
}

@article{eisner2025autistic,
  author  = {Eisner, Daniel C.},
  title   = {Autistic Employees: Workplace Barriers and Solutions for Positive Employment Outcomes},
  journal = {The Journal of Healthcare Ethics \& Administration},
  volume  = {11},
  number  = {4},
  pages   = {59--82},
  year    = {2025},
  doi     = {10.22461/jhea.9.7171},
  url     = {https://doi.org/10.22461/jhea.9.7171}
}

@inproceedings{haroon2025neurobridge,
  author    = {Haroon, Rukhshan and Wigdor, Kyle and Yang, Katie and Toumanios, Nicole and Crehan, Eileen T. and Dogar, Fahad},
  title     = {NeuroBridge: Using Generative {AI} to Bridge Cross-neurotype Communication Differences through Neurotypical Perspective-taking},
  booktitle = {Proceedings of the 27th International ACM SIGACCESS Conference on Computers and Accessibility},
  series    = {ASSETS '25},
  location  = {Denver, CO, USA},
  year      = {2025},
  numpages  = {19},
  pages     = {1--19},
  publisher = {Association for Computing Machinery},
  address   = {New York, NY, USA},
  doi       = {10.1145/3663547.3746337},
  url       = {https://doi.org/10.1145/3663547.3746337}
}

@article{hedley2017employment,
  author  = {Hedley, Darren and Uljarevi{\'c}, Mirko and Cameron, Lauren and Halder, Santoshi and Richdale, Amanda and Dissanayake, Cheryl},
  title   = {Employment Programmes and Interventions Targeting Adults with Autism Spectrum Disorder: A Systematic Review of the Literature},
  journal = {Autism},
  volume  = {21},
  number  = {8},
  pages   = {929--941},
  year    = {2017},
  doi     = {10.1177/1362361316661855},
  url     = {https://doi.org/10.1177/1362361316661855}
}

@article{heinze2025workplace,
  author  = {Heinze, C{\'e}cile},
  title   = {Workplace Accommodations and Employment Outcomes Among Employees With Autism: A Systematic Review},
  journal = {Cureus},
  volume  = {17},
  number  = {12},
  pages   = {e99353},
  year    = {2025},
  doi     = {10.7759/cureus.99353},
  url     = {https://doi.org/10.7759/cureus.99353}
}

@article{kagohara2013using,
  author  = {Kagohara, Debora M. and van der Meer, Larah and Ramdoss, Sathiyaprakash and O'Reilly, Mark F. and Lancioni, Giulio E. and Davis, Tonya N. and Rispoli, Mandy and Lang, Russell and Marschik, Peter B. and Sutherland, Dean and Green, Vanessa A. and Sigafoos, Jeff},
  title   = {Using {iPods}\textregistered{} and {iPads}\textregistered{} in Teaching Programs for Individuals with Developmental Disabilities: A Systematic Review},
  journal = {Research in Developmental Disabilities},
  volume  = {34},
  number  = {1},
  pages   = {147--156},
  year    = {2013},
  doi     = {10.1016/j.ridd.2012.07.027},
  url     = {https://doi.org/10.1016/j.ridd.2012.07.027}
}

@article{kenny2026tell,
  author  = {Kenny, Neil and Doyle, Alison and O'Neill, Claire and Doyle, Jessica K. and O'Kelly, Jane and Earley, Fiona and Neilson, Stuart},
  title   = {``Tell Me What My Job Is'': A Qualitative Exploration of the Experiences of Autistic Academic Staff Working in Higher Education in {Ireland}},
  journal = {Autism},
  volume  = {30},
  number  = {7},
  pages   = {1724--1737},
  year    = {2026},
  doi     = {10.1177/13623613261440799},
  url     = {https://doi.org/10.1177/13623613261440799}
}

@inproceedings{ma2026chatgpt,
  author    = {Ma, Renkai and Zhang, Ben Zefeng and Chen, Chen and Yang, Fan and Huang, Xiaoshan and Wu, Haolun and Li, Lingyao},
  title     = {``{I} Use {ChatGPT} to Humanize My Words'': Affordances and Risks of {ChatGPT} to Autistic Users},
  booktitle = {Proceedings of the 2026 ACM Interactive Health Conference},
  series    = {IH '26},
  year      = {2026},
  numpages  = {8},
  pages     = {1--8},
  location  = {Porto, Portugal},
  publisher = {Association for Computing Machinery},
  address   = {New York, NY, USA},
  doi       = {10.1145/3786579.3804943},
  url       = {https://doi.org/10.1145/3786579.3804943}
}

@article{milton2012ontological,
  author  = {Milton, Damian E. M.},
  title   = {On the Ontological Status of Autism: The `Double Empathy Problem'},
  journal = {Disability \& Society},
  volume  = {27},
  number  = {6},
  pages   = {883--887},
  year    = {2012},
  doi     = {10.1080/09687599.2012.710008},
  url     = {https://doi.org/10.1080/09687599.2012.710008}
}

@article{milton2022ten,
  author  = {Milton, Damian and Gurbuz, Emine and L{\'o}pez, Beatriz},
  title   = {The `Double Empathy Problem': Ten Years On},
  journal = {Autism},
  volume  = {26},
  number  = {8},
  pages   = {1901--1903},
  year    = {2022},
  doi     = {10.1177/13623613221129123},
  url     = {https://doi.org/10.1177/13623613221129123}
}

@article{mueller2026employment,
  author  = {Mueller, Sophia B. and Libster, Natalie and Sullivan, Virginia and DaWalt, Leann Smith and Taylor, Julie Lounds},
  title   = {Associations Between Aspects of Employment and Psychological Health Among Autistic Adults},
  journal = {Autism in Adulthood},
  year    = {2026},
  pages   = {25739581261418877},
  doi     = {10.1177/25739581261418877},
  url     = {https://doi.org/10.1177/25739581261418877},
  note    = {Advance online publication}
}

@article{myles2001understanding,
  author  = {Myles, Brenda Smith and Simpson, Richard L.},
  title   = {Understanding the Hidden Curriculum: An Essential Social Skill for Children and Youth with {Asperger} Syndrome},
  journal = {Intervention in School and Clinic},
  volume  = {36},
  number  = {5},
  pages   = {279--286},
  year    = {2001},
  doi     = {10.1177/105345120103600504},
  url     = {https://doi.org/10.1177/105345120103600504}
}

@book{myles2004hidden,
  author    = {Myles, Brenda Smith and Trautman, Melissa L. and Schelvan, Ronda L.},
  title     = {The Hidden Curriculum: Practical Solutions for Understanding Unstated Rules in Social Situations},
  publisher = {Autism Asperger Publishing Company},
  address   = {Shawnee Mission, KS},
  year      = {2004},
  isbn      = {9781931282604}
}

@article{nevala2019effectiveness,
  author  = {Nevala, Nina and Pehkonen, Irmeli and Teittinen, Antti and Vesala, Hannu T. and P{\"o}rtfors, Pia and Anttila, Heidi},
  title   = {The Effectiveness of Rehabilitation Interventions on the Employment and Functioning of People with Intellectual Disabilities: A Systematic Review},
  journal = {Journal of Occupational Rehabilitation},
  volume  = {29},
  number  = {4},
  pages   = {773--802},
  year    = {2019},
  doi     = {10.1007/s10926-019-09837-2},
  url     = {https://doi.org/10.1007/s10926-019-09837-2}
}

@misc{ocali2019strategies,
  author       = {{Ohio Center for Autism and Low Incidence (OCALI)}},
  title        = {Strategies for Employment Success for {ASD} -- Executive Function},
  howpublished = {Employment and the Youth with Autism Spectrum Disorders guide},
  year         = {2019},
  url          = {https://www.ocali.org/project/2019-Employment-Guide/page/executive-function}
}

@misc{ocalindtips,
  author       = {{Ohio Center for Autism and Low Incidence (OCALI)}},
  title        = {Daily Tips to Address the Hidden Curriculum},
  year         = {n.d.},
  url          = {https://www.ocali.org/project/hidden_curriculum_month}
}

@article{raymaker2023cog,
  author  = {Raymaker, Dora M. and Sharer, Mirah and Maslak, Joelle and Powers, Laurie E. and McDonald, Katherine E. and Kapp, Steven K. and Moura, Ian and Wallington, Anna Furra and Nicolaidis, Christina},
  title   = {``{[I]} Don't Wanna Just Be Like a Cog in the Machine'': Narratives of Autism and Skilled Employment},
  journal = {Autism},
  volume  = {27},
  number  = {1},
  pages   = {65--75},
  year    = {2023},
  doi     = {10.1177/13623613221080813},
  url     = {https://doi.org/10.1177/13623613221080813}
}

@article{rizzo1970role,
  author  = {Rizzo, John R. and House, Robert J. and Lirtzman, Sidney I.},
  title   = {Role Conflict and Ambiguity in Complex Organizations},
  journal = {Administrative Science Quarterly},
  volume  = {15},
  number  = {2},
  pages   = {150--163},
  year    = {1970},
  doi     = {10.2307/2391486},
  url     = {https://doi.org/10.2307/2391486}
}

@article{shattuck2012postsecondary,
  author  = {Shattuck, Paul T. and Narendorf, Sarah Carter and Cooper, Benjamin and Sterzing, Paul R. and Wagner, Mary and Taylor, Julie Lounds},
  title   = {Postsecondary Education and Employment Among Youth With an Autism Spectrum Disorder},
  journal = {Pediatrics},
  volume  = {129},
  number  = {6},
  pages   = {1042--1049},
  year    = {2012},
  doi     = {10.1542/peds.2011-2864},
  url     = {https://doi.org/10.1542/peds.2011-2864}
}

@article{snellrood2020stakeholder,
  author  = {Snell-Rood, Claire and Ruble, Lisa and Kleinert, Harold and McGrew, John H. and Adams, Medina and Rodgers, Alexis and Odom, Jaye and Wong, Wing Hang and Yu, Yue},
  title   = {Stakeholder Perspectives on Transition Planning, Implementation, and Outcomes for Students with Autism Spectrum Disorder},
  journal = {Autism},
  volume  = {24},
  number  = {5},
  pages   = {1164--1176},
  year    = {2020},
  doi     = {10.1177/1362361319894827},
  url     = {https://doi.org/10.1177/1362361319894827}
}

@article{szechy2024autism,
  author  = {Szechy, Kathryn A. and Turk, Pamela D. and O'Donnell, Lisa A.},
  title   = {Autism and Employment Challenges: The Double Empathy Problem and Perceptions of an Autistic Employee in the Workplace},
  journal = {Autism in Adulthood},
  volume  = {6},
  number  = {2},
  pages   = {205--217},
  year    = {2024},
  doi     = {10.1089/aut.2023.0046},
  url     = {https://doi.org/10.1089/aut.2023.0046}
}

@article{thorpe2024lived,
  author  = {Thorpe, David and McKinlay, Marisa and Richards, James and Sang, Kate and Stewart, Mary E.},
  title   = {The Lived Experience of Autistic Adults in Employment: A Systematic Search and Synthesis},
  journal = {Autism in Adulthood},
  volume  = {6},
  number  = {4},
  pages   = {495--509},
  year    = {2024},
  doi     = {10.1089/aut.2022.0114},
  url     = {https://doi.org/10.1089/aut.2022.0114}
}

@article{auger2013speculative,
  author  = {Auger, James},
  title   = {Speculative Design: Crafting the Speculation},
  journal = {Digital Creativity},
  volume  = {24},
  number  = {1},
  pages   = {11--35},
  year    = {2013},
  doi     = {10.1080/14626268.2013.767276},
  url     = {https://doi.org/10.1080/14626268.2013.767276}
}

@misc{choi2025imagining,
  author        = {Choi, Jeanne and Choi, Dasom and Jeong, Sejun and Hong, Hwajung and Seering, Joseph},
  title         = {How Far {I}'ll Go: Imagining Futures of Conversational {AI} with People with Visual Impairments Through Design Fiction},
  year          = {2025},
  eprint        = {2510.12268},
  archivePrefix = {arXiv},
  primaryClass  = {cs.HC},
  url           = {https://arxiv.org/abs/2510.12268}
}

@book{dunneraby2013speculative,
  author    = {Dunne, Anthony and Raby, Fiona},
  title     = {Speculative Everything: Design, Fiction, and Social Dreaming},
  publisher = {MIT Press},
  address   = {Cambridge, MA},
  year      = {2013},
  isbn      = {9780262019842}
}

@inproceedings{elsden2017speculative,
  author    = {Elsden, Chris and Chatting, David and Durrant, Abigail C. and Garbett, Andrew and Nissen, Bettina and Vines, John and Kirk, David S.},
  title     = {On Speculative Enactments},
  booktitle = {Proceedings of the 2017 CHI Conference on Human Factors in Computing Systems},
  series    = {CHI '17},
  year      = {2017},
  pages     = {5386--5399},
  numpages  = {14},
  publisher = {Association for Computing Machinery},
  address   = {New York, NY, USA},
  doi       = {10.1145/3025453.3025503},
  url       = {https://doi.org/10.1145/3025453.3025503}
}

@inproceedings{engstrom2025llm,
  author    = {Engstr{\o}m, Maria Padilla and L{\o}vlie, Anders Sundnes},
  title     = {Using a Large Language Model as Design Material for an Interactive Museum Installation},
  booktitle = {Companion Publication of the 2025 ACM Designing Interactive Systems Conference},
  series    = {DIS '25 Companion},
  year      = {2025},
  pages     = {386--390},
  numpages  = {5},
  publisher = {Association for Computing Machinery},
  address   = {New York, NY, USA},
  doi       = {10.1145/3715668.3736359},
  url       = {https://doi.org/10.1145/3715668.3736359}
}

@inproceedings{hutchinson2003technology,
  author    = {Hutchinson, Hilary and Mackay, Wendy and Westerlund, Bosse and Bederson, Benjamin B. and Druin, Allison and Plaisant, Catherine and Beaudouin-Lafon, Michel and Conversy, St{\'e}phane and Evans, Helen and Hansen, Heiko and Roussel, Nicolas and Eiderb{\"a}ck, Bj{\"o}rn and Lindquist, Sinna and Sundblad, Yngve},
  title     = {Technology Probes: Inspiring Design for and with Families},
  booktitle = {Proceedings of the SIGCHI Conference on Human Factors in Computing Systems},
  series    = {CHI '03},
  year      = {2003},
  pages     = {17--24},
  numpages  = {8},
  publisher = {Association for Computing Machinery},
  address   = {New York, NY, USA},
  doi       = {10.1145/642611.642616},
  url       = {https://doi.org/10.1145/642611.642616}
}

@inproceedings{kozubaev2020expanding,
  author    = {Kozubaev, Sandjar and Elsden, Chris and Howell, Noura and Juul S{\o}ndergaard, Marie Louise and Merrill, Nick and Schulte, Britta and Wong, Richmond Y.},
  title     = {Expanding Modes of Reflection in Design Futuring},
  booktitle = {Proceedings of the 2020 CHI Conference on Human Factors in Computing Systems},
  series    = {CHI '20},
  year      = {2020},
  pages     = {1--15},
  numpages  = {15},
  publisher = {Association for Computing Machinery},
  address   = {New York, NY, USA},
  doi       = {10.1145/3313831.3376526},
  url       = {https://doi.org/10.1145/3313831.3376526}
}

@inproceedings{lc2023speculative,
  author    = {LC, Ray and Tang, Yuying},
  title     = {Speculative Design with Generative {AI}: Applying Stable Diffusion and {ChatGPT} to Imagining Climate Change Futures},
  booktitle = {Proceedings of the 11th International Conference on Digital and Interactive Arts},
  series    = {ARTECH '23},
  year      = {2023},
  pages     = {1--8},
  numpages  = {8},
  publisher = {Association for Computing Machinery},
  address   = {New York, NY, USA},
  doi       = {10.1145/3632776.3632827},
  url       = {https://doi.org/10.1145/3632776.3632827}
}

@article{braun2006using,
  author  = {Braun, Virginia and Clarke, Victoria},
  title   = {Using Thematic Analysis in Psychology},
  journal = {Qualitative Research in Psychology},
  volume  = {3},
  number  = {2},
  pages   = {77--101},
  year    = {2006},
  doi     = {10.1191/1478088706qp063oa},
  url     = {https://doi.org/10.1191/1478088706qp063oa}
}

@article{braun2019reflecting,
  author  = {Braun, Virginia and Clarke, Victoria},
  title   = {Reflecting on Reflexive Thematic Analysis},
  journal = {Qualitative Research in Sport, Exercise and Health},
  volume  = {11},
  number  = {4},
  pages   = {589--597},
  year    = {2019},
  doi     = {10.1080/2159676X.2019.1628806},
  url     = {https://doi.org/10.1080/2159676X.2019.1628806}
}

@book{verbeek2005what,
  author    = {Verbeek, Peter-Paul},
  title     = {What Things Do: Philosophical Reflections on Technology, Agency, and Design},
  publisher = {Penn State University Press},
  address   = {University Park, PA},
  year      = {2005},
  doi       = {10.5325/j.ctv14gp4w7},
  url       = {https://doi.org/10.5325/j.ctv14gp4w7}
}

@article{pezzimenti2024termination,
  author  = {Pezzimenti, Florencia and Durrani, Eman and Zheng, Shuting and Adams, Ryan E. and Bishop, Somer L. and Taylor, Julie Lounds},
  title   = {Perspectives on Employer-Initiated Terminations among Young Adults on the Autism Spectrum},
  journal = {Journal of Autism and Developmental Disorders},
  year    = {2024},
  volume  = {54},
  number  = {4},
  pages   = {1332--1343},
  doi     = {10.1007/s10803-022-05884-6},
  url     = {https://doi.org/10.1007/s10803-022-05884-6}
}

@article{taylor2015longitudinal,
  author  = {Taylor, Julie Lounds and Henninger, Natalie A. and Mailick, Marsha R.},
  title   = {Longitudinal Patterns of Employment and Postsecondary Education for Adults with Autism and Average-Range {IQ}},
  journal = {Autism},
  year    = {2015},
  volume  = {19},
  number  = {7},
  pages   = {785--793},
  doi     = {10.1177/1362361315585643},
  url     = {https://doi.org/10.1177/1362361315585643}
}

@article{scott2019factors,
  author  = {Scott, Melissa and Milbourn, Ben and Falkmer, Marita and Black, Melissa and B{\"o}lte, Sven and Halladay, Alycia and Lerner, Matthew and Taylor, Julie Lounds and Girdler, Sonya},
  title   = {Factors Impacting Employment for People with Autism Spectrum Disorder: A Scoping Review},
  journal = {Autism},
  year    = {2019},
  volume  = {23},
  number  = {4},
  pages   = {869--901},
  doi     = {10.1177/1362361318787789},
  url     = {https://doi.org/10.1177/1362361318787789}
}

@article{ellenkamp2016work,
  author  = {Ellenkamp, Joke J. H. and Brouwers, Evelien P. M. and Embregts, Petri J. C. M. and Joosen, Margot C. W. and van Weeghel, Jaap},
  title   = {Work Environment-Related Factors in Obtaining and Maintaining Work in a Competitive Employment Setting for Employees with Intellectual Disabilities: A Systematic Review},
  journal = {Journal of Occupational Rehabilitation},
  year    = {2016},
  volume  = {26},
  number  = {1},
  pages   = {56--69},
  doi     = {10.1007/s10926-015-9586-1},
  url     = {https://doi.org/10.1007/s10926-015-9586-1}
}

@inproceedings{gaver2003ambiguity,
  author    = {Gaver, William W. and Beaver, Jacob and Benford, Steve},
  title     = {Ambiguity as a Resource for Design},
  booktitle = {Proceedings of the SIGCHI Conference on Human Factors in Computing Systems},
  series    = {CHI '03},
  year      = {2003},
  pages     = {233--240},
  numpages  = {8},
  publisher = {Association for Computing Machinery},
  address   = {New York, NY, USA},
  doi       = {10.1145/642611.642653},
  url       = {https://doi.org/10.1145/642611.642653}
}

@inproceedings{blythe2016antisolutionist,
  author    = {Blythe, Mark and Andersen, Kristina and Clarke, Rachel and Wright, Peter},
  title     = {Anti-Solutionist Strategies: Seriously Silly Design Fiction},
  booktitle = {Proceedings of the 2016 CHI Conference on Human Factors in Computing Systems},
  series    = {CHI '16},
  year      = {2016},
  pages     = {4968--4978},
  numpages  = {11},
  publisher = {Association for Computing Machinery},
  address   = {New York, NY, USA},
  doi       = {10.1145/2858036.2858482},
  url       = {https://doi.org/10.1145/2858036.2858482}
}

@article{shaw2025prevalence,
  author  = {Shaw, Kelly A. and Williams, Susan and Patrick, Mary E. and others},
  title   = {Prevalence and Early Identification of Autism Spectrum Disorder among Children Aged 4 and 8 Years --- Autism and Developmental Disabilities Monitoring Network, 16 Sites, United States, 2022},
  journal = {MMWR Surveillance Summaries},
  year    = {2025},
  volume  = {74},
  number  = {2},
  pages   = {1--22},
  doi     = {10.15585/mmwr.ss7402a1},
  url     = {https://doi.org/10.15585/mmwr.ss7402a1}
}

@article{roux2013postsecondary,
  author  = {Roux, Anne M. and Shattuck, Paul T. and Cooper, Benjamin P. and Anderson, Kristy A. and Wagner, Mary and Narendorf, Sarah C.},
  title   = {Postsecondary Employment Experiences among Young Adults with an Autism Spectrum Disorder},
  journal = {Journal of the American Academy of Child and Adolescent Psychiatry},
  year    = {2013},
  volume  = {52},
  number  = {9},
  pages   = {931--939},
  doi     = {10.1016/j.jaac.2013.05.019},
  url     = {https://doi.org/10.1016/j.jaac.2013.05.019}
}

@techreport{nci2026employment,
  author      = {{National Core Indicators}},
  title       = {{NCI-IDD} In-Person Survey 2024--25 National Report: Employment},
  institution = {Human Services Research Institute and National Association of State Directors of Developmental Disabilities Services},
  year        = {2026},
  url         = {https://idd.nationalcoreindicators.org/survey-reports-insights/},
  note        = {Retrieved September 10, 2026}
}

@article{fillary2006social,
  author  = {Fillary, Rose and Pernice, Regina},
  title   = {Social Inclusion in Workplaces Where People with Intellectual Disabilities Are Employed: Implications for Supported Employment Professionals},
  journal = {International Journal of Rehabilitation Research},
  year    = {2006},
  volume  = {29},
  number  = {1},
  pages   = {31--36},
  doi     = {10.1097/01.mrr.0000185952.87304.63},
  url     = {https://doi.org/10.1097/01.mrr.0000185952.87304.63}
}

@article{scheef2019salient,
  author  = {Scheef, Andrew R. and Walker, Zachary M. and Barrio, Brenda L.},
  title   = {Salient Employability Skills for Youth with Intellectual and Developmental Disabilities in {Singapore}: The Perspectives of Job Developers},
  journal = {International Journal of Developmental Disabilities},
  year    = {2019},
  volume  = {65},
  number  = {1},
  pages   = {1--9},
  doi     = {10.1080/20473869.2017.1335479},
  url     = {https://doi.org/10.1080/20473869.2017.1335479}
}

@article{cavanagh2021subtle,
  author  = {Cavanagh, Jillian and Meacham, Hannah and Pariona-Cabrera, Patricia and Bartram, Timothy},
  title   = {Subtle Workplace Discrimination Inhibiting Workers with Intellectual Disability from Thriving at the Workplace},
  journal = {Personnel Review},
  year    = {2021},
  volume  = {50},
  number  = {7/8},
  pages   = {1739--1756},
  doi     = {10.1108/PR-10-2021-0723},
  url     = {https://doi.org/10.1108/PR-10-2021-0723}
}

@article{petnerarrey2016facilitating,
  author  = {Petner-Arrey, Jami and Howell-Moneta, Angela and Lysaght, Rosemary},
  title   = {Facilitating Employment Opportunities for Adults with Intellectual and Developmental Disability through Parents and Social Networks},
  journal = {Disability and Rehabilitation},
  year    = {2016},
  volume  = {38},
  number  = {8},
  pages   = {789--795},
  doi     = {10.3109/09638288.2015.1061605},
  url     = {https://doi.org/10.3109/09638288.2015.1061605}
}

@article{balasuriya2022support,
  author  = {Balasuriya, Saminda Sundeepa and Sitbon, Laurianne and Brereton, Margot},
  title   = {A Support Worker Perspective on Use of New Technologies by People with Intellectual Disabilities},
  journal = {ACM Transactions on Accessible Computing},
  year    = {2022},
  volume  = {15},
  number  = {3},
  pages   = {1--21},
  numpages = {21},
  doi     = {10.1145/3523058},
  url     = {https://doi.org/10.1145/3523058}
}

@article{kruithof2024parents,
  author  = {Kruithof, Kasper and Hoogsteyns, Maartje and Zaal-Schuller, Ilse and Huisman, Sylvia and Willems, Dick and Nieuwenhuijse, Appolonia},
  title   = {Parents' Tacit Knowledge of Their Child with Profound Intellectual and Multiple Disabilities: A Qualitative Study},
  journal = {Journal of Intellectual \& Developmental Disability},
  year    = {2024},
  volume  = {49},
  number  = {4},
  pages   = {415--424},
  doi     = {10.3109/13668250.2024.2336084},
  url     = {https://doi.org/10.3109/13668250.2024.2336084}
}

@article{koch2015proxy,
  author  = {Koch, Andrea D. and Vogel, Anke and Becker, Thomas and Salize, Hans-Joachim and Voss, Elke and Werner, Amelie and Arnold, Katrin and Sch{\"u}tzwohl, Matthias},
  title   = {Proxy and Self-Reported Quality of Life in Adults with Intellectual Disabilities: Impact of Psychiatric Symptoms, Problem Behaviour, Psychotropic Medication and Unmet Needs},
  journal = {Research in Developmental Disabilities},
  year    = {2015},
  volume  = {45--46},
  pages   = {136--146},
  doi     = {10.1016/j.ridd.2015.07.022},
  url     = {https://doi.org/10.1016/j.ridd.2015.07.022}
}

@article{berastegui2021reporters,
  author   = {Berastegui, Ana and Santos, Angela and Su{\'a}rez, Graciela},
  title    = {The Reporters' Agreement in Assessing the Quality of Life of Young People with Intellectual Disabilities},
  journal  = {Research in Developmental Disabilities},
  year     = {2021},
  volume   = {116},
  articleno = {104026},
  numpages = {10},
  doi      = {10.1016/j.ridd.2021.104026},
  url      = {https://doi.org/10.1016/j.ridd.2021.104026}
}

\appendix
\section{Interview Protocol}
\label{app:protocol}

Table~\ref{tab:protocol} reports the core questions and probes used in the semi-structured interviews and the AI co-design activity described in Section~\ref{subsec:data-collection}. Interviewers asked the core questions of every participant and selected one or two probes per domain rather than working through all of them. We did not show participants the four ambiguity labels, and we introduced each domain through an open question about the person they support.

\begingroup
\setlength{\tabcolsep}{4pt}
\scriptsize
\begin{longtable}{@{}>{\raggedright\arraybackslash}p{0.16\textwidth}
                    >{\raggedright\arraybackslash}p{0.40\textwidth}
                    >{\raggedright\arraybackslash}p{0.38\textwidth}@{}}
\caption{Semi-structured interview protocol. Core questions were asked of every participant; probes were used selectively as the conversation required.}
\label{tab:protocol}\\
\toprule
\textbf{Interview section} & \textbf{Core questions} & \textbf{Probes} \\
\midrule
\endfirsthead
\multicolumn{3}{@{}l}{\footnotesize\textit{Table~\ref{tab:protocol} continued from previous page.}}\\
\toprule
\textbf{Interview section} & \textbf{Core questions} & \textbf{Probes} \\
\midrule
\endhead
\midrule
\multicolumn{3}{r@{}}{\footnotesize\textit{Continued on next page.}}\\
\endfoot
\bottomrule
\endlastfoot

\textbf{Role and context} &
``Can you tell me a little about your role and how you are connected to autistic or IDD individuals in workplace or work-like settings?'' \newline
``Without sharing any names or identifying details, could you describe a person you support who has experience in a job, internship, or volunteer position?'' \newline
``What AI chatbot or service, if any, do you typically use in your daily or professional life?'' &
``Are you mainly a family caregiver, a support professional, a job coach, or something else?'' \newline
``About how long have you been in this role?'' \newline
``What kind of work do they do, and who do they interact with most, for example supervisors, coworkers, or customers?'' \\
\addlinespace[2pt]

\textbf{Open definition of ambiguity} &
``When you think about that person's experiences at work, what kinds of situations feel unclear, confusing, or hard to read to them?'' \newline
``If you had to put it in your own words, what does `ambiguity' at work mean in their world?'' &
``This could be about what people say, how people act, or rules and expectations, anything that makes them think `I'm not sure what this means.' Can you think of any examples?'' \\
\addlinespace[2pt]

\textbf{Instructional ambiguity} &
``Can you tell me about a time when the person you support was unsure what they were expected or instructed to do at work?'' &
``What made that situation feel unclear to them?'' \newline
``How did they try to figure it out?'' \newline
``How did other people around them respond?'' \\
\addlinespace[2pt]

\textbf{Social-norm ambiguity} &
``Thinking about social interactions at work, can you describe a situation where it was hard for the person to tell what the `right' way to act or respond was?'' \newline
``Are there any `unwritten rules' at work that the person you support has had to figure out the hard way?'' &
``This might be about tone of voice, jokes, small talk, or body language. What happened?'' \newline
``How did they interpret the situation, and how did others seem to interpret it?'' \newline
``Can you think of a time when they did something that seemed perfectly okay to them, but others saw it differently?'' \\
\addlinespace[2pt]

\textbf{Evaluative ambiguity} &
``Can you tell me about a time when it wasn't clear to the person how others felt about their behavior or performance at work?'' \newline
``Have there been moments when they were unsure whether something they did would lead to consequences, positive or negative?'' &
``What kinds of signals did they notice from supervisors or coworkers?'' \newline
``Did those signals ever seem to send different messages at the same time?'' \newline
``What happened afterward, and did the outcome match what they expected?'' \\
\addlinespace[2pt]

\textbf{Navigation ambiguity} &
``When situations at work feel confusing for the person, what options do they feel they have for getting help or advice, and do they actually know how to access those options?'' \newline
``Are there any workplace rules, policies, or support services that have felt unclear or hard to navigate, either for you or for them?'' &
``Are they ever unsure who it is okay to talk to, for example a supervisor, a coworker, a job coach, or a family member?'' \newline
``Can you describe a time when they wanted support but weren't sure how to ask, or who to ask?'' \newline
``What made those rules or services feel confusing, and how did you eventually figure them out?'' \\
\addlinespace[2pt]

\textbf{AI activity: selecting a scenario} &
``From everything we've talked about today, can you pick one specific situation at work that felt especially unclear or tricky for the person you support?'' &
``What about that situation makes it a good example of the kind of ambiguity we've been discussing?'' \\
\addlinespace[2pt]

\textbf{AI activity: formulating a prompt} &
``Imagine you had a very helpful assistant sitting next to you who is great with AI tools. If you were going to ask that assistant for help with this specific situation, what would you ask them to do?'' &
Interviewers offered two examples of prompt phrasing without supplying content: ``Help me come up with three different ways X could respond to Y,'' and ``Explain this situation in simple steps so X can understand what's happening.'' \\
\addlinespace[2pt]

\textbf{AI activity: reviewing the output} &
``Looking at what the AI gave us, what parts feel like they could actually be helpful for the person you support in that situation, and why?'' \newline
``Are there any parts that feel confusing, unrealistic, or not safe to use?'' &
``Overall, how satisfied are you with this response, thinking about the person you support, and what would you want to change?'' \\
\addlinespace[2pt]

\textbf{AI activity: refining the prompt} &
``If you could change your original request to make the AI's answer more useful, what would you add or take out?'' \newline
``Does this version feel closer to what you were looking for, and what's still missing?'' &
Interviewers suggested refinements only when a participant was unsure, such as making the explanation shorter and more concrete, using simpler language, or including an example script. \\
\addlinespace[2pt]

\textbf{Closing} &
``Based on everything we did today, in what kinds of social situations at work do you think AI tools like this could genuinely be helpful for autistic or IDD workers?'' \newline
``Are there any situations where you would not want them to rely on AI for support, and why?'' &
``How could this solution potentially help in tackling other forms of ambiguity we discussed today?'' \newline
``Is there anything important about confusing or unclear situations at work, and how you try to help navigate them, that we haven't covered today but you think we should understand?'' \\

\end{longtable}
\endgroup

\section{Thematic Codebook and Researcher Positionality}
\label{app:codebook}
Tables~\ref{tab:codebook-rq1} and~\ref{tab:codebook-rq2} report the codebook produced by the analysis described in Section~\ref{subsec:data-analysis}. Representative quotations are short excerpts from caregiver interviews, lightly punctuated for readability, with ellipses marking omitted text. Employer names, workplace details, and other identifying information have been removed.

\begingroup
\setlength{\tabcolsep}{4pt}
\scriptsize
\begin{longtable}{@{}>{\raggedright\arraybackslash}p{0.20\textwidth}
                    >{\raggedright\arraybackslash}p{0.24\textwidth}
                    >{\raggedright\arraybackslash}p{0.50\textwidth}@{}}
\caption{RQ1 codebook of workplace ambiguity. Five primary themes and 13 sub-themes, developed from 153 initial codes.}
\label{tab:codebook-rq1}\\
\toprule
\textbf{Primary theme} & \textbf{Sub-theme} & \textbf{Representative quotation} \\
\midrule
\endfirsthead
\multicolumn{3}{@{}l}{\footnotesize\textit{Table~\ref{tab:codebook-rq1} continued from previous page.}}\\
\toprule
\textbf{Primary theme} & \textbf{Sub-theme} & \textbf{Representative quotation} \\
\midrule
\endhead
\midrule
\multicolumn{3}{r@{}}{\footnotesize\textit{Continued on next page.}}\\
\endfoot
\bottomrule
\endlastfoot
\textbf{The Hidden Curriculum of Workplace Social Norms} & Navigating Multi-Layered Workplace Relationships & ``The social pieces, the relationships with other employees --- he was putting way too much stock in his relationship with this supervisor, not realizing that it was actually not his supervisor, but the one he worked with that was having an issue with him." [P1]'' \\
 & Reputational Costs of Strict Norm Compliance & ``At the beginning of his job he gets a lot of compliments for being such a rule follower, and then can you guess what happens after that? He starts policing other people on those rules." [P1]'' \\
 & Context-Dependent Unwritten Social Norms & ``You can smoke, and cuss, and spit, and wear blue jeans when you're a construction worker. None of those things are gonna be successful in a banking environment." [P1]'' \\
\addlinespace[2pt]
\textbf{Difficulty in Evaluating Performance Expectations} & Adapting to Changing Task Instructions & ``If the task changes --- first we were sorting this part, doing these things, taking this one thing and sorting it into these different containers, and now we have to do a different kind of sort, a different task. Pretty routine-oriented in her daily life --- a new routine in the middle of the workday is [hard]." [P2]'' \\
 & Perfectionism as a Barrier to Seeking Feedback & ``I work with individuals on the autism spectrum that are perfectionists, so they feel like once they've been told what they're supposed to do, they're just supposed to execute it to perfection." [P1]'' \\
 & Misreading Feedback as Optional & ``The feedback was just kind of like a nice-to-have instead of a --- this is going to make or break my job if I don't apply this feedback." [P1]'' \\
 & Managing Emotional Responses in Public Work Spaces & ``He was feeling like he was gonna have a meltdown, getting overwhelmed and whatnot, so he kept trying to tamp down instead of taking a break and going and engaging in some self-regulation." [P1]'' \\
\addlinespace[2pt]
\textbf{Misjudgement of Social Boundaries with Non-Disabled Peers} & Difficulty Calibrating Social Engagement with the Public & ``Not really even how, but, you know, enough's enough. Like, get back to work, talking, even if they're working and talking --- knowing when it's time to." [P2]'' \\
\addlinespace[2pt]
\textbf{Opacity of Formal Workplace Structures} & Confusion About Workplace Roles and Procedures & ``This kid is letting me know that he's not gonna make it to work, but he needs to be letting his supervisor know. There's too many cooks in the kitchen when you have a job coach involved." [P1]'' \\
 & Accommodation Knowledge Gaps & ``We don't do a great job at the school level of educating individuals about accommodations. We have all these beautiful documents that have their accommodations listed, but then we don't say, hey, you really benefit from extended time --- this is how you could request it." [P1]'' \\
\addlinespace[2pt]
\textbf{Barriers to Accessing Workplace Support} & Institutional Pushback Against Accommodation Requests & ``He reached out to his supervisor and included me in the email, and introduced me as a job coach. The supervisor said he wasn't allowed to talk to me, took me off the email, and never talked to me again." [P1]'' \\
 & Peer Stigma Around Visible Accommodation & ``One of his coworkers was kind of like, why is he getting all this extra help? And you have to explain, it's not extra help, it just doesn't come as naturally to him as it does to you." [P3]'' \\
 & Lack of Structured Support Throughout Employment & ``I've always suggested to the employees that I support requesting weekly touch bases when they first start, and then moving to less frequent as they get more comfortable. But a lot of employers don't --- some people have never had a one-on-one with their boss." [P1]'' \\
\end{longtable}
\endgroup

\begingroup
\setlength{\tabcolsep}{4pt}
\scriptsize
\begin{longtable}{@{}>{\raggedright\arraybackslash}p{0.17\textwidth}
                    >{\raggedright\arraybackslash}p{0.21\textwidth}
                    >{\raggedright\arraybackslash}p{0.20\textwidth}
                    >{\raggedright\arraybackslash}p{0.36\textwidth}@{}}
\caption{RQ2 codebook of technology support. Three primary themes and 10 sub-themes, developed from 54 initial codes. The third column lists the RQ1 ambiguity themes that each support was proposed to address, following the participant's own account of the situation.}
\label{tab:codebook-rq2}\\
\toprule
\textbf{Primary theme} & \textbf{Sub-theme} & \textbf{Ambiguity addressed (RQ1)} & \textbf{Representative quotation} \\
\midrule
\endfirsthead
\multicolumn{4}{@{}l}{\footnotesize\textit{Table~\ref{tab:codebook-rq2} continued from previous page.}}\\
\toprule
\textbf{Primary theme} & \textbf{Sub-theme} & \textbf{Ambiguity addressed (RQ1)} & \textbf{Representative quotation} \\
\midrule
\endhead
\midrule
\multicolumn{4}{r@{}}{\footnotesize\textit{Continued on next page.}}\\
\endfoot
\bottomrule
\endlastfoot
\textbf{Ambient Real-Time Behavioral Redirection} & Conversation-Duration Monitoring via Wearable & Hidden curriculum; Performance expectations; Social boundaries & ``I would want it to be able to gently, kindly remind them to stay on task --- probably audio cues. If a customer goes through checkout in 3 minutes average, your conversation probably shouldn't be 3 minutes or something.'' [P2] \\
 & Proximity Detection for Physical Boundary Management & Performance expectations; Social boundaries & ``Some sort of device that would alert them that they're too close to somebody. Something that would detect, like, you're too close, you're too close.'' [P3] \\
 & Adaptive Sensory Environment Control & Performance expectations & ``I would like to better control the sensory parts of the environment that trigger meltdowns --- which could be noise, hot temperatures, or other people in the environment.'' [P5] \\
 & Task Prompting Tied to Performance Goals & Hidden curriculum; Performance expectations; Formal structures; Accessing support & ``It could be a task reminder in your calendar --- that's how I do it. Make sure to update my appraisal for whatever goals I'm working on. An Apple Watch is a great one as well.'' [P1] \\
\addlinespace[2pt]
\textbf{Technology-Mediated Feedback Processing} & Feedback Translation from Supervisor to Actionable Steps & Performance expectations; Accessing support & ``I think it would probably be taking feedback from supervision with your supervisor, and then applying it, and then reporting back.'' [P1] \\
 & Social Cue Interpretation via App-Based Guidance & Social boundaries & ``Some sort of app for social skills --- how do I know when someone's joking? Show me facial expressions, stereotypical facial expressions that mean they like talking to me, or some reciprocal stuff like that.'' [P3] \\
 & Employer-Facing Communication Guidance & Hidden curriculum; Performance expectations; Social boundaries; Formal structures; Accessing support & ``On the employer side, if they had something that said, okay, my employee has autism --- if I say this, this, and this, what does that mean to him? How might they perceive what I'm telling them to do? When I say do X, Y, Z, are they going to understand that it's okay to do Y and then X and then Z?'' [P4] \\
 & Technology for Work Interest Discovery & Hidden curriculum; Performance expectations; Social boundaries; Accessing support & ``Could you put them in a virtual [reality]? What's it like to be in a factory, in a kitchen, in a restaurant, what are all those sounds? Because until you go experience it, you don't know what will be appealing to someone until they get a chance to see the inside of a place they've never been.'' [P5] \\
\addlinespace[2pt]
\textbf{Physiologically Informed Socio-Emotional Guidance} & Biofeedback as a Trigger for Social Behavioral Guidance & Performance expectations; Social boundaries & ``Notice your heartbeat is higher than usual --- you might need to take a break before you talk to your supervisor. That would be super awesome --- you might need to go self-regulate because it feels like you're a little discombobulated right now.'' [P1] \\
 & Confidence Prompting at Points of Task Transition & Performance expectations & ``If they started a new task and were trained, they could have a reminder at some point --- if you need help, ask your supervisor, remember you've got this. Just something to, like, a little pep talk buddy.'' [P2] \\
\end{longtable}
\endgroup

\section{Researcher Positionality}
\label{sec:researcher-positionality}
Our interpretation of caregivers' accounts rests on our positionality, including our social backgrounds, intellectual histories, and lived experience. One author has long worked with autistic and IDD communities, which grounded our reading of the episodes participants described in how employment support operates day to day. Another author is neurodivergent, and their perspective informed how we read episodes in which behavior that coworkers misread was, from the worker's side, an ordinary response to an underspecified situation. A third author has family experience with autism and IDD, which kept the caregiving labor in our data concrete. The remaining authors research mental health and behavioral intervention, and they reviewed our findings. Their expertise shaped how we coded accounts of anxiety, overwhelm, and emotional regulation at work, which we treated as responses to environments that withheld information rather than as individual deficits. Because caregivers, rather than workers, supplied these accounts, we returned to this distinction throughout analysis and coded what participants observed and did, not what they inferred about workers' internal states. We discussed these standpoints in our weekly meetings and used them to test, rather than to confirm, the themes we were building.


\end{document}